\documentclass[12pt, a4paper, parskip, english]{scrartcl}
\usepackage{babel}
\usepackage{lmodern}
\usepackage[utf8]{inputenc} 
\usepackage[T1]{fontenc}    
\usepackage[%
    backend=biber,
    style=nature,
    url=false,
    doi=false,
    date=year,
    doi=true,
    isbn=false,
]{biblatex}
\usepackage{graphicx}
\graphicspath{ {./figures/} }   
\usepackage{mwe}            
\usepackage{hyperref}       
\usepackage{url}            
\usepackage{booktabs}       
\usepackage{amsfonts}       
\usepackage{microtype}      
\usepackage{physics}        
\usepackage{csquotes}       
\usepackage[format=plain, font=small, labelfont=bf, labelsep=period]{caption}
\usepackage{subcaption}
\usepackage{cleveref}       
\usepackage{xcolor}
\usepackage{authblk}

\title{Entropy-driven cell-decision making predicts fluid-to-solid transition in multicellular systems}
\date{\today}
\author[a,*]{Arnab Barua}
\author[b,\thanks{Contributed equally}]{Simon Syga} 
\author[a]{Pietro Mascheroni} 
\author[c]{Nikos Kavallaris} 
\author[a,d]{Michael Meyer-Hermann}
\author[b]{Andreas Deutsch}
\author[a,b]{Haralampos Hatzikirou}

\affil[a]{Department of Systems Immunology and Braunschweig Integrated Centre of
Systems Biology, Helmholtz Centre for Infection Research, Rebenring 56,
38106 Braunschweig, Germany}
\affil[b]{Technische Univesit\"at Dresden, Center for Information Services and High Performance Computing, N\"othnitzer Stra{\ss}e 46, 01062, Dresden, Germany}
\affil[c]{Department of Mathematical and Physical Sciences, University of Chester, Chester}
\affil[d]{Institute for Biochemistry, Biotechnology and Bioinformatics,Technische Universität Braunschweig, Braunschweig, Germany}
\providecommand{\keywords}[1]
{
  \small	
  \textbf{\textit{Keywords---}} #1
}

\begin{document}
\maketitle
\begin{abstract}
\noindent Cellular decision making allows cells to assume functionally different phenotypes in response to microenvironmental cues, without genetic change. It is an open question, how individual cell decisions influence the dynamics at the tissue level. Here, we study spatio-temporal pattern formation in a population of cells exhibiting phenotypic plasticity, which is a paradigm of cell decision making. We focus on the migration/resting and the migration/proliferation plasticity which underly the epithelial-mesenchymal transition (EMT) and the go or grow dichotomy. We assume that cells change their phenotype in order to minimize their microenvironmental entropy (LEUP: Least microEnvironmental Uncertainty Principle) and study the impact of the LEUP-driven migration/resting and migration/proliferation plasticity on the corresponding multicellular spatio-temporal dynamics with a stochastic cell-based mathematical model for the spatio-temporal dynamics of the cell phenotypes. In the case of the go or rest plasticity,  a corresponding mean-field approximation allows to identify a bistable switching mechanism between a diffusive (fluid) and an epithelial (solid) tissue phase which depends on the sensitivity of the phenotypes to the environment. For the go or grow plasticity, we show the possibility of Turing pattern formation for the "solid" tissue phase and its relation with the parameters of the LEUP-driven cell decisions.

\end{abstract}
\vspace{1cm}
\hspace{0.7cm}\keywords{Cell-decision making; Phenotypic plasticity; Langevin equations; Mean-field theory; Least microEnvironmental Uncertainty Principle (LEUP)}; Fluid-to-solid transition; Spatio-temporal patterns

\section{Introduction}
Cellular decision making allows cells to respond to microenvironmental cues \cite{Bowsher2014}.
For instance, if an external ligand from the microenvironment binds to the cell membrane, it may trigger a chain of biochemical reactions inside the cell \cite{bruce}.
After the microenvironmental information has been processed, the cell may respond by changing its properties implying cell differentiation, proliferation, migration, apoptosis etc.
Thus, cellular decision making depends on intrinsic signal transduction pathways, the genetic cell network \cite{Prochazka2017}, extrinsic stimuli, and molecular noise \cite{Balazsi2011}.
Cell-decision making plays a key role in cell fate determination and the maintenance of phenotypic plasticity.
In the former, cells irreversibly acquire new fates by following a hierarchical lineage, where pluripotent stem cells in a proper microenvironment (stem cell niche) differentiate into e.~g. bone, muscle, epithelial and further specialised cells.
On the other hand, phenotypic plasticity allows a reversible adaptation to different
microenvironmental stimuli \cite{Price2003}. While cellular decision making is
well-studied for single cells, it is an open question  how individual cell decisions influence the dynamics at the tissue level.

In this paper, we analyse the implications of phenotypic plasticity on spatio-temporal pattern formation in the cell population and focus on the migration/resting and migration/proliferation plasticity. The migration/resting type of plasticity is observed during the epithelial-mesenchymal transition (EMT/MET)\cite{kalluri, Radisky4325, THIERY2003740} while the migration/proliferation plasticity characterizes the go-or-grow (GoG) dichotomy found in different types of cancers, such as glioblastoma or breast cancer, but also during normal development \cite{Hatzikirou2010a,Jerby2012,Hoek2008,Kohrman2017}.
Epithelial cells are immotile, connected by strong adhesion bonds, exhibit regular shapes and show strong apico-basal cell polarity. Epithelial and mesenchymal phenotypes can be interpreted as representing two migration modes, resting and motile, respectively. On the other hand, GoG refers to the mutual exclusion of cell migration and proliferation, leading to the existence of proliferative/immotile and nonproliferative/mobile cell phenotypes. We assume that a Least microEnvironmental Uncertainty Princple (LEUP) dictates the phenotypic switching in both types of phenotypic plastic behaviours. The LEUP principle has been introduced in \cite{Hatzikirou2018} and assumes that cells change their phenotypes in order to minimize their local microenvironmental entropy. Formulated in the language of statistical physics and Bayesian statistics, cells try to optimize their phenotypic prior to reduce the joint entropy of their own phenotype and microenvironmental variables. Similar ideas using information theoretic measures have been proposed also in seminal works by W. Bialek \cite{Bialek2012a}. 

The manuscript is structured as follows: In the next section (2), we define a LEUP-driven cell decision model based on the migration/resting plasticity. In Section 3, we develop a stochastic individual-based model (IBM) for moving and resting cells where cells can change their phenotype according to the LEUP-driven cell decision model and show the formation of aggregates of resting cells in corresponding simulations. In section 4, we derive a macroscopic mean-field approximation of our microscopic model which allows to explain the formation of aggregates in the microscopic model. In section 5, we analyze the stability of the steady states  and the pattern formation potential of the macroscopic go-or grow model. Finally, in section 6, we discuss the biological implications of our results in terms of multicellular growth and pattern control.

\section{A minimal model of LEUP-driven cell migration plasticity}
Here, we define a mathematical model for LEUP-based switching between
moving and resting phenotypes.
We assume that cells can sense their microenvironment and can change their own phenotype $X_{i}$ on a domain $\mathcal{L}\subset \mathbb{R}^{2}$  accordingly, where $X_i = 0, 1$ corresponds to the resting and migrating state, respectively.
We define the microenvironment of cell $i$ as the number $N_{i}^{0} $ of cells having phenotype $X_{i} = 0$ and by the number $N_{i}^{1} $ of cells having phenotype $X_{i} = 1$. 
By assuming a maximum cell capacity $N$ of the microenvironment 
, we can define
\begin{equation}\label{def:N}
    N=N_{i}^{1}+N_{i}^{0}+ N_{i}^{\phi},
\end{equation}
where $N_{i}^{\phi}$ are the free spaces/slots.
The total number of cells $N_{T}$ is defined as the sum of cells having phenotype $\left(X_{i} = 1\right)$ and cells having phenotype $\left(X_{i} = 0\right)$. So, 
\begin{equation}
   N_{T} = N_{i}^{1}+ N_{i}^{0}.
\end{equation}
Cells will select their phenotype in a Bayesian fashion, i.e.
\begin{equation}
P\left(X_i\mid Y_i\right)=\frac{P\left(Y_i\mid X_i\right)P\left(X_i\right)}{P\left(Y_i\right)},
\end{equation}
where $P\left(Y_i\mid X_i\right)$ can be interpreted as the probability that the cell perceives all other cells in its surroundings, and $P\left(X_i\right)$ is the prior probability distribution of the cell's phenotypes.
The former models the microenvironmental sensing the intrinsic programming of cellular phenotype.
However, sensing other cells and evaluating $P\left(Y_i\mid X_i\right)$ entails an energy cost.
It is reasonable to assume that the cell will try to optimize its prior $P\left(X_i\right)$ for the sake of energetic frugality.

 Here we define $Y_{i}$ as the extrinsic random variable of the $i$-th cell, where $Y_{i}=N_{i}^{X_{i}}\mid N_T,X_{i}=0,1 $ represents the number of cells of phenotype $X_{i}$ given the total local number of cells  $N_T$.
 The conditional probability of having $N_{i}^{1}$ number of cells present in the microenvironment  follows a binomial distribution $\left(\boldsymbol{B}\right)$ (for the derivation details see S.I.):  
\begin{equation}
    P\left(N_{i}^{1}\mid N_T\right)= P\left(Y_{i}=N_{i}^{1}\mid N_T \right)=\boldsymbol{B}\left(N_T, p_1 \right),
    \label{bin}
\end{equation}
 where $p_1$ is the probability of $N_{i}^{1}$ number of cells having  phenotype $X_{i}$ = 1 out of $N_T$ cells:
\begin{equation}
    p_1 = \frac{ N_{i}^{1}}{N_T}.
\end{equation}
Using statistical mechanical arguments \cite{Hatzikirou2018}, the problem of finding the probability of the cell's phenotype is equivalent to finding the prior that minimizes the entropy of a cell's phenotype and its surroundings. 
Let $S\left(X_i,Y_i\right)$ be the entropy of the cell-surroundings, $S\left(X_i\right)$ the internal entropy of the cell, and $S\left(Y_i\mid X_i\right)$ the entropy of the information sensed by the cell.
Then, the entropies are connected by the relation $S\left(X_i,Y_i\right)=S\left(X_i\right)+S\left(Y_i\mid X_i\right)$.
The optimization problem is finding $P\left(X_i\right)$ that minimizes $S\left(X_i,Y_i\right)$, while making sure that $P\left(X_i\right)$ is normalized; in other words we impose
\begin{equation}
\begin{split}
  \frac{\delta}{\delta P\left(X_i\right)} & \left\{S\left(X_i\right)+\beta\left[\sum P\left(X_i \right)S\left(Y_i\mid X_i\right)-\bar{S}\left(Y_i\mid X_i\right)\right]\right. 
 \\& \left.-\lambda\left[\sum P\left(X_i\right)-1\right]\right\}=0,
 \end{split}
 \label{functional}
\end{equation}
where $\frac{\delta}{\delta P\left(X_i\right)}$ is the functional derivative, $\bar{S}\left(Y_i\mid X_i\right)$ is the expected ensemble statistics \cite{Hatzikirou2018}, and $\lambda$ and $\beta$ are Lagrange multipliers. Taking into account the relations among entropies, eqn.~(\ref{functional}) yields
\begin{equation}\label{pg}
P\left(X_i \right) = \frac{e^{\beta S\left(Y_i\mid X_i \right)}}{Z}=\frac{1}{1+e^{\beta\Delta S}},
\end{equation}
where $Z$ is the normalization factor
\begin{equation}
   Z =\sum_{\alpha=0,1} e^{\beta S\left(Y_i\mid X_i = \alpha\right)}= e^{\beta S(Y_i|X_i = 0)}+e^{\beta S(Y_i|X_i = 1)}.
\end{equation}
Please note that the parameter $\beta$ has a biological interpretation, since it quantifies the intensity with which a cell senses and complies to the microenvironment. To estimate the entropy we use the binomial distribution of the resting and migrating cells.
Given $N_T$ cells in the environment and assuming the frequency of cells at a moving state $p_1 = \frac{N^{1}_{i}}{N_T}$, the probability for $N_{i}^{1}$ moving cells is 
\begin{equation}
    P\left(N^{1}_{i}\mid N_T\right) = \binom{N_T}{N^{1}_{i}} p_1^{N_{i}^{1}} (1-p_1)^{N_{T}-N^{1}_{i}},
\end{equation}
and the entropy of the binomial distribution in the last equation is 
\begin{equation}
    S(N^{1}_{i}, p_1) = -\sum_{N^{1}_{i}=0}^{N_{T}} P(N^{1}_{i}, p_1) \ln P(N^{1}_{i}, p_1).
\end{equation}
When there are no cells in the microenvironment the entropy $S(N^{1}_{i}, p_1)$ is 0.
To facilitate the evaluation of the entropy distribution of the microenvironment, we have used a Gaussian distribution, which approximates the binomial distribution (for $N_T>5$), i.e. 
\begin{equation}
    P( N^{1}_{i}, p_{1}) \rightarrow{\boldsymbol{N}(N_{T}p_1,N_Tp_1(1-p_1))},
\end{equation}
for the microenvironment \cite{box}.
Now we can evaluate the equilibrium distribution  $P(X_{i}=0)$ as
\begin{equation}
\begin{split}
P\left(X_i = 0\right) &=\frac{e^{\beta S\left(Y_i\mid X_i = 0\right)}}{Z}=\frac{1}{1+e^{\beta\Delta S}} =\\
&=\frac{1}{1+\left[\frac{N_i^{0}\left(N_i^{1}-1\right)}{N_i^{1}\left(N_i^{0}-1\right)}\right]^{\frac{\beta}{2}}} = \frac{1}{1+\left[\frac{\rho_i^{0}\left(\rho_i^{1}-\frac{1}{V}\right)}{\rho_i^{1}\left(\rho_i^{0}-\frac{1}{V}\right)}\right]^{\frac{\beta}{2}}},
 \label{pgeneral}
 \end{split}
\end{equation}
with the cell densities $\rho_i^{0,1} := \frac{N_i^{0,1}}{V}$ (in the SI Fig.~(\ref{restprob}), we show the dependency of $P\left(X_i = 0\right)$ on resting density $\rho_0$ and $\beta$). Accordingly, the probability of a $i$-th cell's phenotype $P(X_{i}=1)$ reads
\begin{equation}
\begin{split}
P\left(X_i = 1\right) &= \frac{e^{\beta S\left(Y_i\mid X_i = 1\right)}}{Z}=1-\frac{1}{1+e^{\beta\Delta S}}=\\
&=1-\frac{1}{1+\left[\frac{N_i^{0}\left(N_i^{1}-1\right)}{N_i^{1}\left(N_i^{0}-1\right)}\right]^{\frac{\beta}{2}}} = \frac{\left[\frac{\rho_i^{0}\left(\rho_i^{1}-\frac{1}{V}\right)}{\rho_i^{1}\left(\rho_i^{0}-\frac{1}{V}\right)}\right]^{\frac{\beta}{2}}}{1+\left[\frac{\rho_i^{0}\left(\rho_i^{1}-\frac{1}{V}\right)}{\rho_i^{1}\left(\rho_i^{0}-\frac{1}{V}\right)}\right]^{\frac{\beta}{2}}}.
 \label{pgeneral2}
 \end{split}
\end{equation}
Please note that, the difference between the microenvironmental entropy of the $i$-th cell is
\begin{equation}
    \Delta S=S(Y_i|X_{i} =1)- S(Y_i| X_{i} = 0)=\frac{1}{2}\ln{\left[\frac{N_i^{0}\left(N_i^{1}-1\right)}{N_i^{1}\left(N_i^{0}-1\right)}\right]},
    \label{difference}
\end{equation}
$N_i^{0}$ and $N_i^{1} $ are the number of cell of phenotype ($X_{i}$ = 0) 
and phenotype ($X_{i}$ = 1) in the microenvironment of $i$-th cell,  respectively (See S.I. for details). 
\begin{figure}
 \centering
 \begin{subfigure}[b]{0.45\linewidth}
\includegraphics[width=\linewidth]{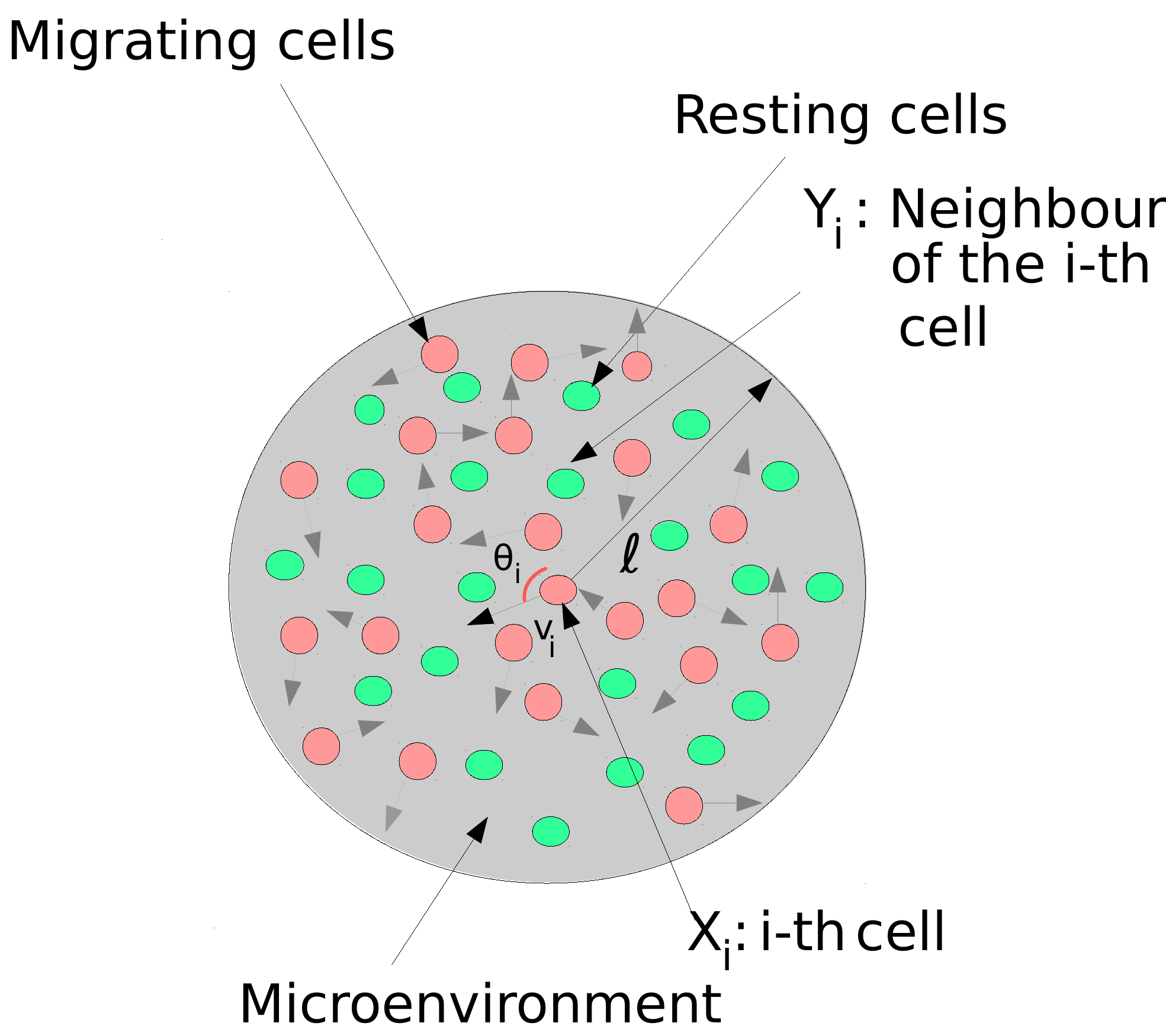}
\caption{}\label{scheme1}
\end{subfigure}
~
\begin{subfigure}[b]{0.45\linewidth}
\includegraphics[width=\linewidth]{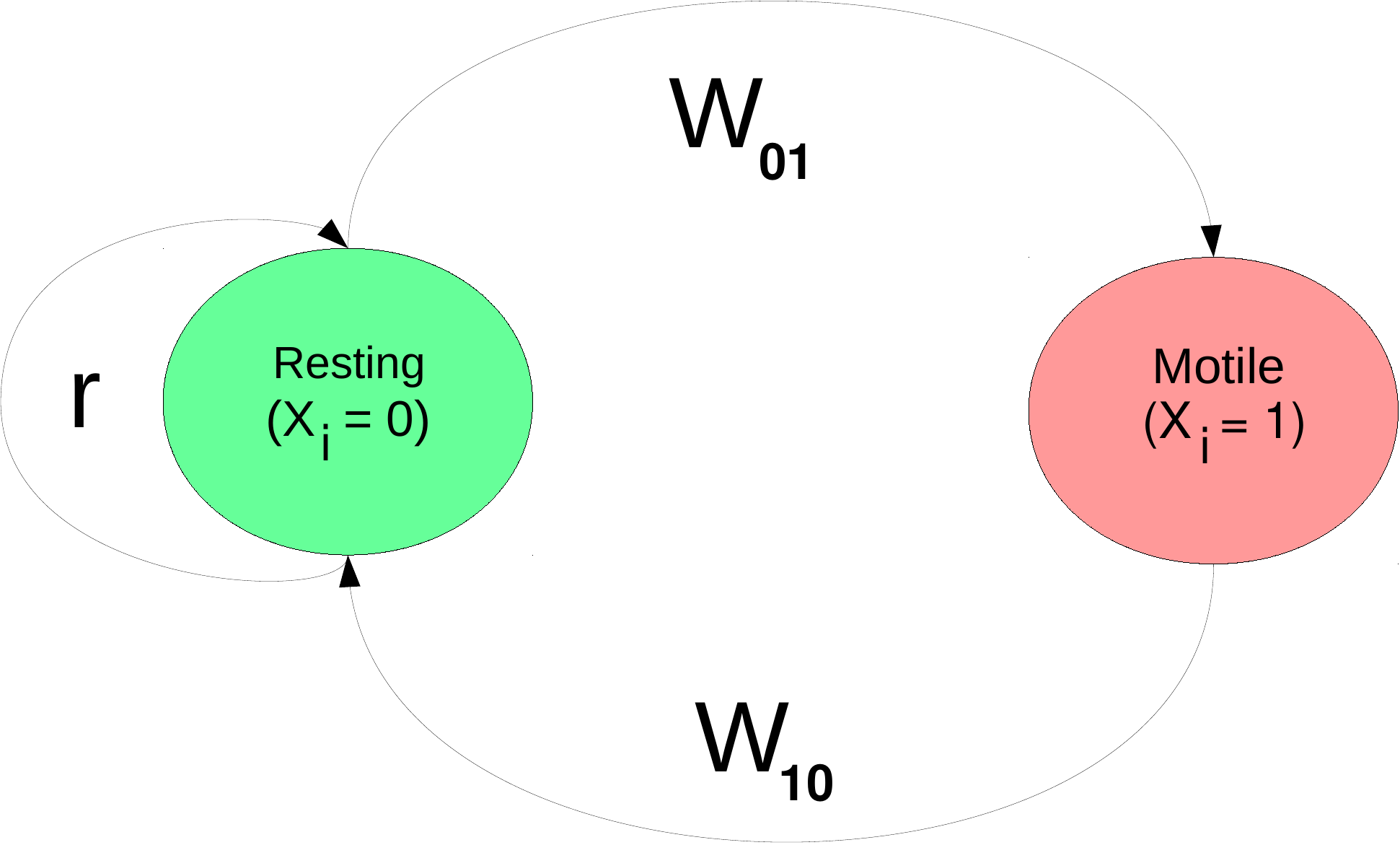}
\caption{}\label{schem2}
\end{subfigure}
 \caption{Schematic diagram of (\textbf{a}) the  microenvironment of a motile cell, where $\ell$ is the sensing radius. The sensed number of cells is proportional to the volume of the sensed microenvironment, i.e. $N\propto \ell^{d}$, where d is the dimension. (\textbf{b}) Transitions in the \enquote{Go or Grow} model \cite{Hatzikirou2010a}. The switching between motile  $\left(X_i = 1\right)$ and resting  $\left(X_i = 0\right)$ phenotype is shown, where the transtion probabilities are defined  as $W_{10}$ and $W_{01}$. The proliferation rate  is defined by $r$.}
 \label{sche}
\end{figure}


Fig.~(\ref{scheme1}), shows a sketch  of our model. In short, we make the following assumptions
\\$\left(A1\right)\hspace{0.5cm}$ The microenvironment is defined by the numbers of two phenotypes only.
\\$\left(A2\right)\hspace{0.5cm}$ We have assumed a binomial distribution for the occurrence of the two phenotypes.
\\$\left(A3\right)\hspace{0.5cm}$ Cells are making decisions at a fast time scale. This justifies to assume an equilibrium
distribution for the different phenotypes.

In this section, we have defined a minimal LEUP-driven cell decision making model and derived the corresponding phenotypic steady states. Based on this, we subsequently develop a cell-based model to understand the resulting multicelluar spatiotemporal dynamics.

\section{An individual-based model (IBM) for LEUP-driven cell migration }
We define a discrete stochastic, spatio-temporal IBM that incorporates the phenotypic switch dynamics according to LEUP.
To this end, we model the movement of single cells with Langevin equations.
Langevin equations are well-suited to model cell-cell interactions and cell migration \cite{Hakim_2017, Schienbein1993}. 

Our Langevin's equation is defined on a domain $\mathcal{L}\subset \mathbb{R}^{2}$ with periodic boundary conditions.
We define an interaction radius $\ell\in\mathbb{R}$ around the $i$-th cell at position $\vb{x_i}\in \mathcal{L}$.
The expected  interaction volume  is $V\propto\ell^{2}$.
The time evolution of our model is defined by the following rules:
\begin{enumerate}
    \item[(R1)] Cells change their phenotype by sensing their microenvironment within the interaction radius $\ell$ according to LEUP.
     \item[(R2)] Moving cells change their orientations randomly (random walk).
    \item[(R3)] Once cells become migratory they move with a constant speed $\bar{v}$. 
    \end{enumerate}
    The above are translated to the following Langevin's equations
\begin{equation}
 \begin{split}
  & \frac{\mathrm{d}\mathbf{x_i}}{\mathrm{d}t} = \bar{v} \mathbf{v_i} \qty(\theta_i ), \\
  & \frac{\mathrm{d} \theta_i}{\mathrm{d}t} = \frac{1}{\bar{v}}\xi_{i}^\theta(t),\\
  & \frac{\mathrm{d} p_i}{\mathrm{d}t} = -\frac{1}{\tau}(p_i-p_i^{eq}),\\
   & v_i= 
\begin{cases}
    \bar{v},&  \qif X_i = 1,\\
    0,      &  \qif X_i = 0,
\end{cases}
 \end{split}
\end{equation}
where $\vb{v_i} = \qty(\cos \theta_i, \sin \theta_i )^T$ is the direction of movement of cell $i$. 
For the temporal evolution of the probability $p_i$ of the motile state, we use the BGK (Bhatnagar–Gross–Krook) operator technique \cite{bhatt}.
We assume that the probability $p_i$ evolves weakly out of its equilibrium probability $p_i^{eq}$, which is the LEUP steady state probability $P(X_i=1)$ (see eq.~(\ref{pgeneral2})).
In turn, the parameter $\tau$ is the relaxation time towards  the corresponding probability distribution $p_i^{eq}$. 
Here noise is assumed to have a zero-mean, white noise term, which has the statistical properties $\left\langle\xi^{\theta}_i(t)\right\rangle=0$ and $\left\langle\xi^{\theta}_i\left(t_1\right)\xi^{\theta}_j\left(t_2\right)\right\rangle=2D_{\theta}\delta\left(t_1-t_2\right)\delta_{ij}$, where $t_1$ and $t_2$ are two time points, $D_{\theta}\in\mathbb{R}_+$ is the angular diffusion coefficient, $\delta(t)$ is the Dirac delta, and $\delta_{ij}$ is the Kronecker delta.
Parameter $\bar{v}$ is the constant speed of every motile cell. 

\begin{figure}
 \centering
 \begin{subfigure}[b]{0.48\linewidth}
\includegraphics[width=\linewidth]{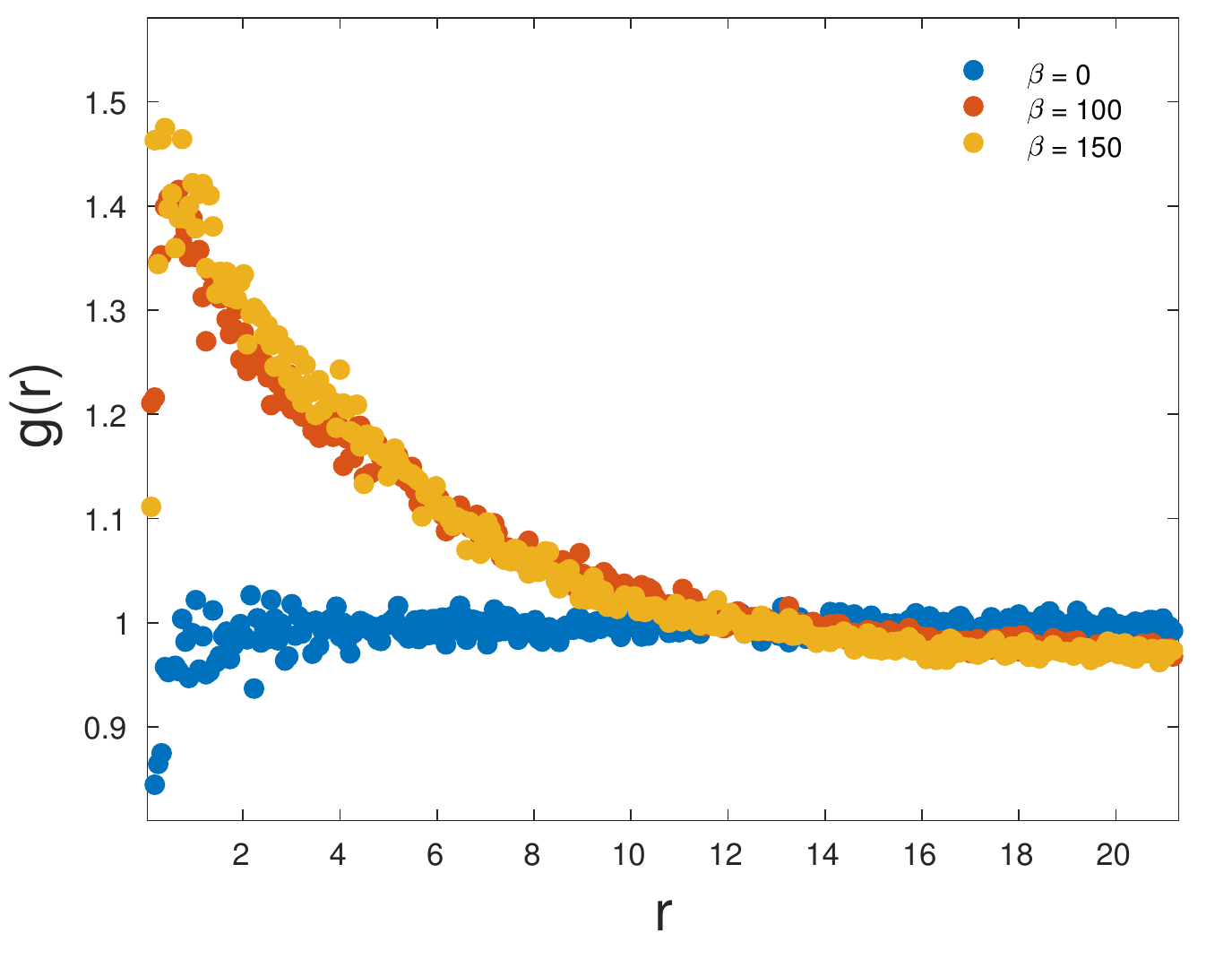}
\caption{}\label{radial1}
\end{subfigure}
~
\begin{subfigure}[b]{0.48\linewidth}
\includegraphics[width=\linewidth]{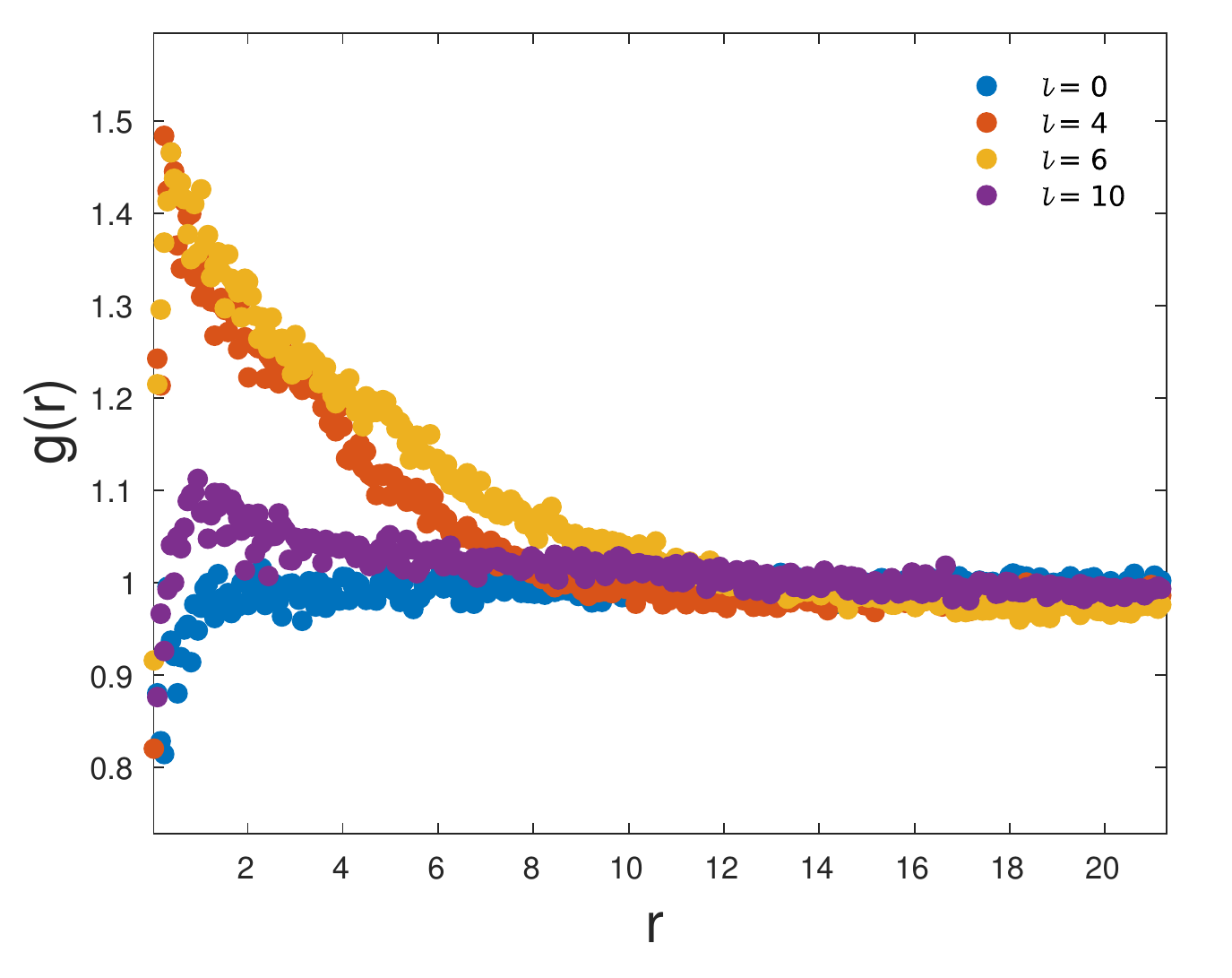}
\caption{}\label{radial2}
\end{subfigure}
~
\begin{subfigure}[b]{0.48\linewidth}
\includegraphics[width=\linewidth]{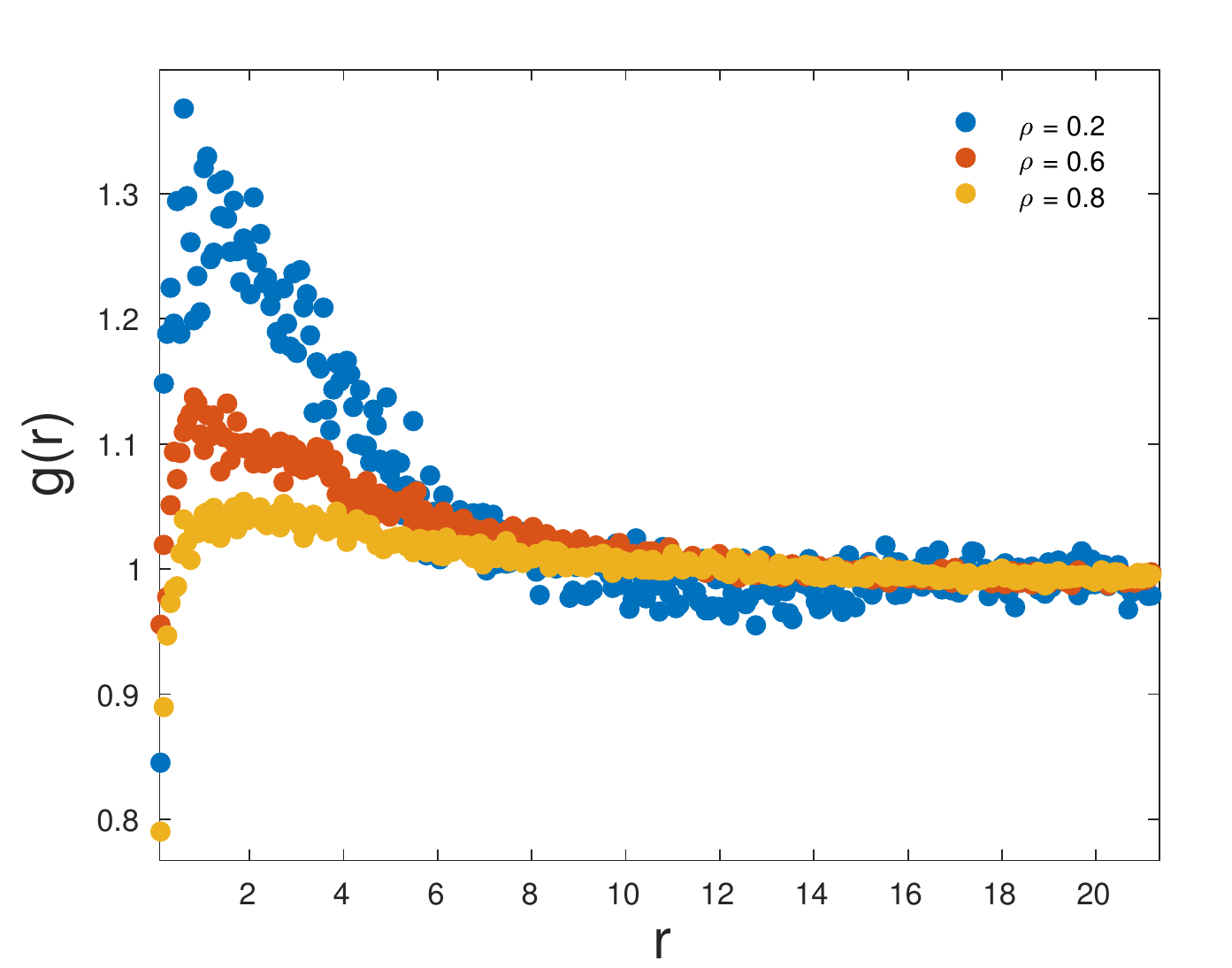}
\caption{}\label{radial3}
\end{subfigure}
~
\begin{subfigure}[b]{0.40\linewidth}
\includegraphics[width=\linewidth]{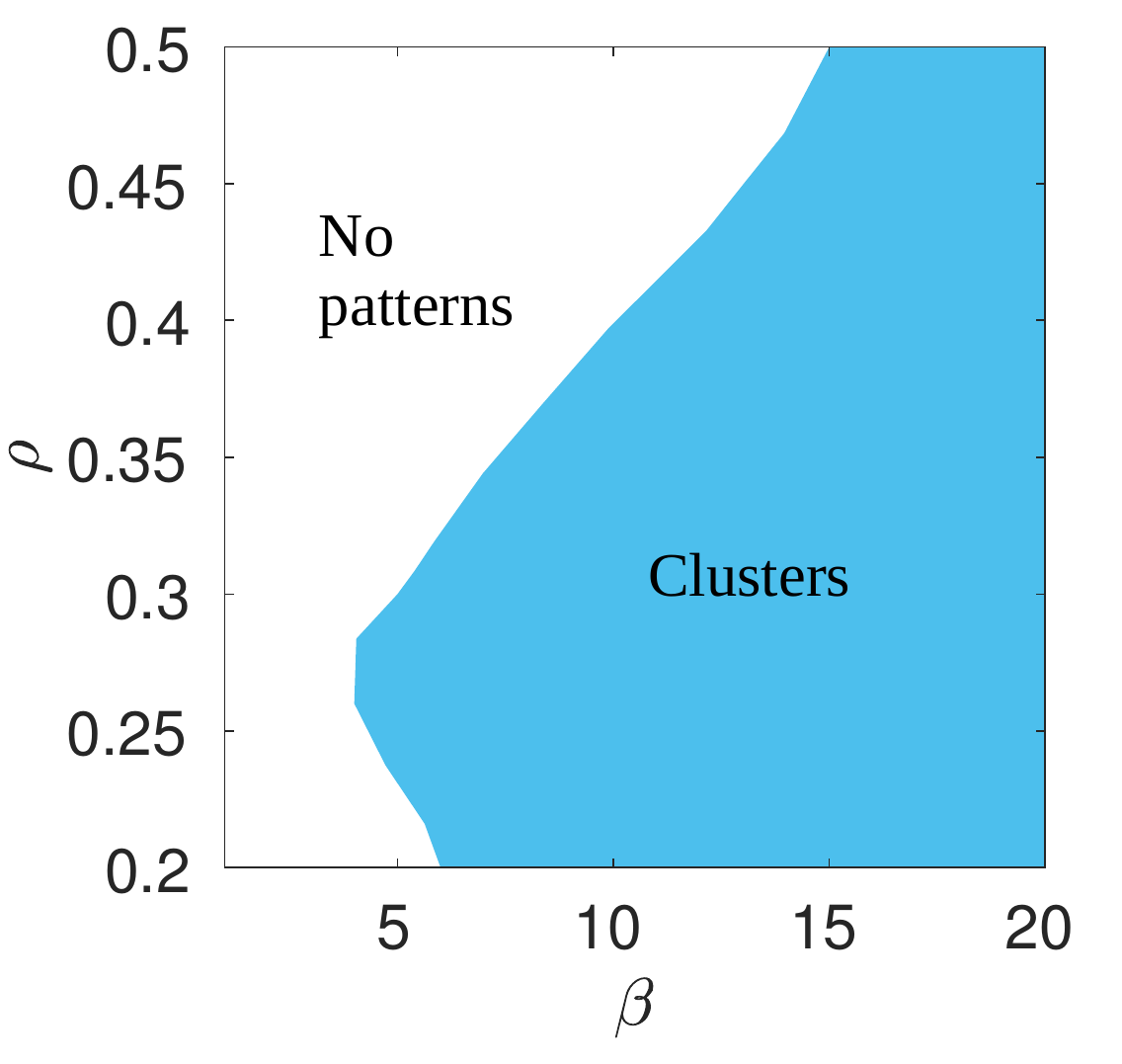}
\caption{}\label{nonspbif}
\end{subfigure}
\caption{Quantification of cell clustering. (\textbf{a}) Radial distribution function  for different values of $\beta$, for a fixed interaction radius $\ell = 6$.
The number of cells was fixed at 1000, corresponding to a mean cell density of 0.4.
(\textbf{b}) Radial distribution function for varying interaction radius $\ell$ at fixed sensitivity $\beta =100$.
There is an optimal sensing radius for a given mean density, so that interaction radii that are too large or too low do not lead to aggregation, indicated by an almost flat radial distribution, see purple and blue lines.
(\textbf{c}) Radial distribution function for different mean densities, at $\beta = 20$ and $\ell = 4$. When the mean density becomes too large, no clustering can be observed.
(\textbf{d}) Phase space diagram for cluster pattern formation.
Here we define clustering when $\max_r g(r)>1.09$ (see text for explaination).
The interaction radius is fixed at $\ell= 4$.
All data points in (\textbf{a - d}) correspond to the mean of 20 independent simulations.}

\end{figure}

We simulate the Langevin model, on a two-dimensional domain, with a varying  interaction radius $\ell$, mean cell density and sensitivity $\beta$. We assume an initial state that is approximately homogeneous in space.
For sufficiently large densities and sensitivities, we observe aggregation patterns, as shown in Figure (\ref{clus}).
To quantify cell clustering, we calculate the radial distribution function for different sensitivities and different interaction radii as shown in Figures (\ref{radial1}) and (\ref{radial2}). The key observations from the simulation study are:
\begin{figure}
  \centering
\includegraphics[scale=0.5]{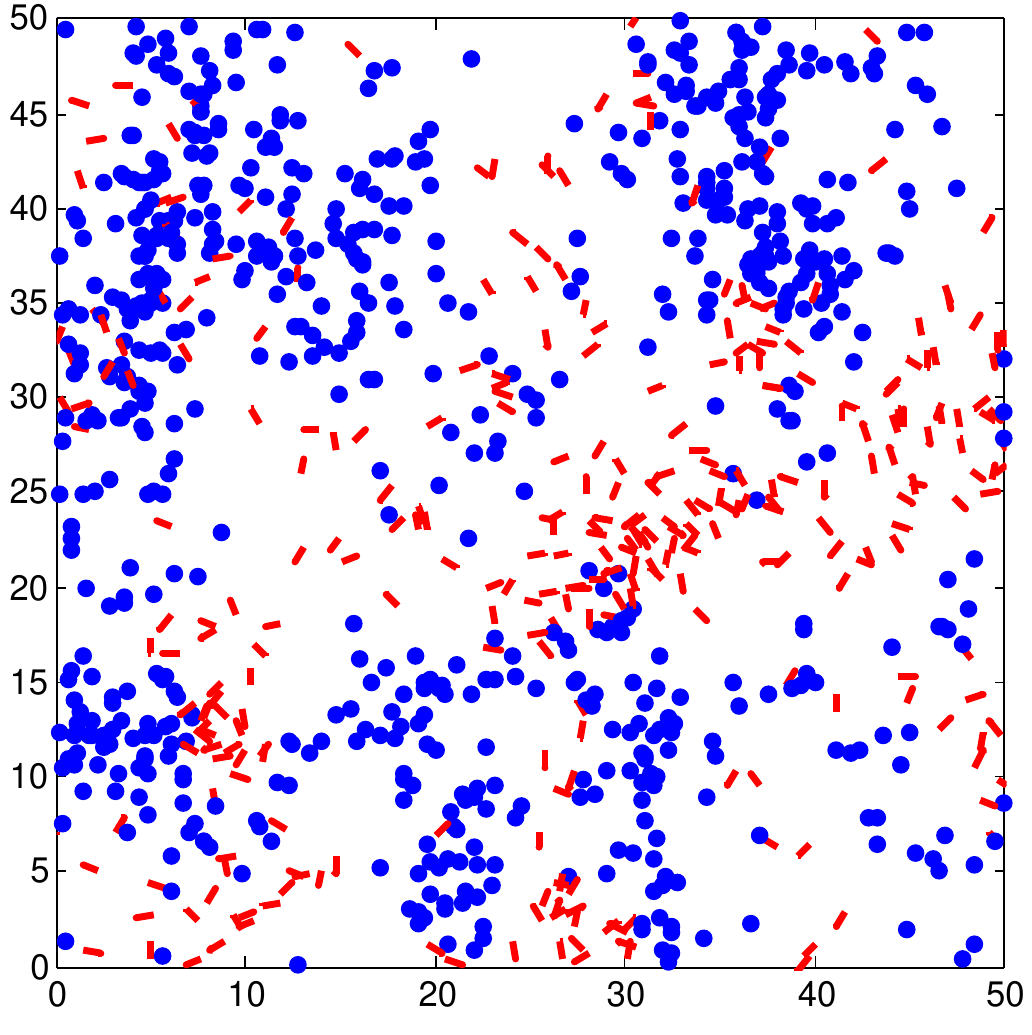}
\caption{Clustering of resting cells in the IBM. Snapshot of an example realization at $t = 100$, where the total number of cells is fixed at 1000 and $\beta=40$.
Resting cells are shown in blue and migrating cells are marked by red arrows indicating their direction of movement.}
\label{clus}
\end{figure}

\begin{enumerate}
    \item There is a critical threshold for the parameter $\beta>0$, where clustering of resting cells occurs (Fig. \ref{radial1}). This is rather expected since a sufficient sensing of the microenvironment is required.
    \item For an intermediate interaction radius $\ell$, we observe cell clusters as shown in Fig. \ref{radial2}.
    Very high $\ell$ corresponds to a large number of sensed cells. This leads to also equal steady state probabilities, as the entropy difference in the microenvironment, associated with the single cell states becomes negligible.
    On the other hand, the lower bound of $\ell$ is expected, since enough sampling size of cellular microenvironment is required to induce aggregation patterns.
	\item In Fig.~\ref{nonspbif}, we show the phase diagram of the system, when parameter $\beta$ and the average density are varied.
	In particular,  we observe that there is a parametric regime where clustering behavior is  emerging.
	Interestingly, high densities do not always imply patterning and require higher values of $\beta$ to support cluster emergence (see Fig. \ref{radial3}).
	Finally, as expected, low densities also reduce the area of the patterning regime for low $\beta$.
\end{enumerate}
\section{Mean-field approximation}
In this section, we derive a continuous approximation of the aforementioned discrete model using a mean-field approach.
The goal is to shed light of the pattern formation mechanisms, i.e. cell clustering, as observed in the microscopic simulations.
The main idea of the mean-field approximation is to replace the description of many-particle interactions by a single particle description based on an average or effective interaction.
Thereby, any multi-particle problem can be replaced by an effective description, that can be stated in the form of an ordinary (ODE) or partial differential equation (PDE).
In order to proceed, we will first treat the switch dynamics and the migration process separately.

\subsection{Mean-field description of the phenotypic switch dynamics}
Let's consider cell $i$ at position $\vb{x_i}$ with states $X_i = 0, 1$.
The cell changes its state in dependence on the local microenvironment according to the LEUP with rates $W_{01},W_{10}$ and the probabilities to be in either states are denoted by $p_{0, 1}$.
We assume that the system is always close to the steady state, so the master equation reads
\begin{equation}
    \dv{}{t} p_0 = W_{10} p_1 - W_{01} p_0 \approx 0,
\end{equation}
and $p_{0, 1}$ are given by \cref{pg}.
By rearranging the terms, we obtain for the switching rates
\begin{equation}
    \frac{W_{10}}{W_{01}} =  \frac{p_0}{p_1}.
\end{equation}
For simplicity, we  set $W_{01} = p_1$ where the transition probability towards motile phenotypes equals to the moving steady state probability $p_1$.
This coincides with the \enquote{detailed balance condition}.
The steady state probabilities $p_{0,1}$ depend on the number of cells $N_i^{0,1}$ in the respective phenotypes in the microenvironment. Cell $i$ at position $\vb{x_i}$ senses 
\begin{equation}\label{eq:n_particles}
    N_i^X(\vb{x_1, \dots, x_i, \dots, x_N}, t) := \sum_{j = 1}^N \xi \qty(\vb{x_j}(t) - \vb{x_i}(t), t) \delta_{X_j, X}  \,, \quad X = 0, 1.
\end{equation}
Here, we sum over all cells $j$, and $\xi \qty(\vb{x_j}(t) - \vb{x_i}(t), t)$ is a Boolean stochastic variable that serves as the sensing function of a cell at $\vb{x_i}$.
It depends on the distance between cells and time, and $\xi = 1$ indicates that the cell $j$ at $\vb{x_j}$ is sensed by cell $i$ at position $\vb{x_i}$.
To match the IBM, we assume that
\begin{equation}\label{eq:heaviside_sensing}
    \xi (\vb{x_j}(t) - \vb{x_i}(t), t) = \Theta\qty(\ell - \abs{\vb{x}_j(t) - \vb{x_i}(t)}), \quad \ell > 0,
\end{equation}
with the Heaviside step function $\Theta(x)$, so that all cells in a ball of radius $\ell$ around $\vb{x_i}$ are sensed.

To proceed, we apply a mean-field approximation to calculate the expected switching rate $\ev{W_{10}( N_i^0, N_i^1) } \approx W_{01} \qty(\ev{N_i^0}, \ev{N_i^1})$.
Note that we dropped the dependence on space and time for better readability.
Let $\phi_i (X, \vb{x}, t)$ denote the probability of finding cell $i$ in a small volume $\dd V$ around $\vb{x}$ at time $t$ with the phenotype $X = 0, 1$. 
Formally we have
\begin{equation}
    \phi_i (X, \vb{x}, t) = \ev{\delta\qty(\vb{x} - \vb{x_i}(t)) \delta_{X, X_i}},
\end{equation}
where the average is the ensemble average, and the total densities of resting/migrating cells are
\begin{align}
    \Phi(\vb{x}, t) &:= \sum_i \phi_i (0, \vb{x}, t), \\
    \Psi (\vb{x}, t) &:= \sum_i \phi_i (1, \vb{x}, t).
\end{align}
For the expected density of sensed cells in the microenvironment around $\vb{x}$ we obtain
\begin{align}\label{eq:nonlocal_cellnumbers}
     \rho_0( \vb{x}, t) := \frac{\ev{N_i^0 ( \vb{x}, t)}}{V} &= \frac{1}{V} \int_{\mathcal{B}_{\ell}(\vb{x})} \Phi(\vb{y}, t) \, \dd V, \\
     \rho_1( \vb{x}, t)  := \frac{\ev{N_i^1 ( \vb{x}, t)}}{V} &= \frac{1}{V} \int_{\mathcal{B}_{\ell}(\vb{x})} \Psi(\vb{y}, t) \, \dd V,
\end{align}
where $V \propto \ell^d$, and where we have already used eq. \cref{eq:heaviside_sensing}.
Consequently, we obtain the approximate switching rate
\begin{equation}
   \bar{W}_{10} = p_0 ( \rho_0,  \rho_1),
\end{equation}
and the switch dynamics is (dropping dependencies on space and time for simplicity)
\begin{equation}
    \pdv{}{t}  \phi_i (0) =   p_0 \phi_i(1) - p_1 \phi_i(0).
\end{equation}
Summing over all cells, we obtain for the total density of resting cells
\begin{equation}\label{eq:ME_switch}
    \pdv{}{t} \Phi =  p_0 \Psi - p_1\Phi.
\end{equation}
Note that this is a non-local, non-linear set of PDEs, which are difficult to treat analytically. However, we can further simplify our analysis by making another approximation, assuming that the sensing radius $\ell$ is much smaller than the  total domain size. Then, we can replace the non-local sensing function $\xi \qty(\vb{x_j}(t) - \vb{x_i}(t), t)$ by a local delta distribution
\begin{equation}
    \xi \qty(\vb{x_j}(t) - \vb{x_i}(t), t) \to \delta \qty(\vb{x_j}(t) - \vb{x_i}(t)).
\end{equation}
In this case, the expected number of sensed cells in the microenvironment around $\vb{x}$ simply becomes the local density of the respective phenotype
\begin{align}\label{eq:local_cellnumbers}
     \rho_0  ( \vb{x}, t):= \ev{N_i^0 ( \vb{x}, t)} &= \frac{\Phi(\vb{x}, t)}{\ell^d}, \\
    \rho_1 ( \vb{x}, t) := \ev{N_i^1 ( \vb{x}, t)} &=\frac{ \Psi(\vb{x}, t)}{\ell^d}.
\end{align}
Finally, the rate $\bar{W}_{10}$ reduces to
\begin{equation}\label{eq:W10_rate}
    \bar{W}_{10}= p_0 ( \rho_0,  \rho_1)= \frac{1}{1+e^{\beta\Delta S}}=\frac{1}{1+\left[\frac{\rho_0\left(\rho_1-\frac{1}{V}\right)}{\rho_1\left(\rho_0-\frac{1}{V}\right)}\right]^{\frac{\beta}{2}}},
\end{equation} 
and the reversed transition probability $\bar{W}_{01}$ is 
\begin{equation}\label{eq:W01_rate}
    \bar{W}_{01}= p_1 ( \rho_0,  \rho_1)=1- \frac{1}{1+e^{\beta\Delta S}}=1-\frac{1}{1+\left[\frac{\rho_0\left(\rho_1-\frac{1}{V}\right)}{\rho_1\left(\rho_0-\frac{1}{V}\right)}\right]^{\frac{\beta}{2}}},
\end{equation}
where the volume is defined as $V=\ell^d$. 
 \subsection{Mean-field description of the cell migration process}
In this section, we derive the macroscopic equation in two spatial dimensions for the motile cell population. As before we can write the corresponding stochastic Langevin's equations as
\begin{equation}
 \begin{split}
& \dv{}{t} \vb{x_i} = \bar{v} \vb{v_i} \qty(\theta_i ), \\
  & \dv{}{t} \theta_i= \frac{1}{\bar{v}}\xi_{i}^\theta(t).\\
  \end{split}
\end{equation}
This process can be considered as a special kind of active Brownian motion.
In this kind of Langevin's equation, the stochastic force creates variations of orientation. According to \cite{Milster2017} we can derive the corresponding Fokker-Planck equation for migrating cells using adiabatic elimination and averaging it over different noise realizations obtaining the following diffusion equation
\begin{equation}
      \frac{\partial \rho_{1}(x,t)}{\partial t}=\frac{\bar{v}^{4}}{2D_{\theta}}\nabla^{2}\rho_{1}(x,t).
\end{equation}
Here we define $D_1$ as diffusion coefficient which is $\frac{\bar{v}^4}{2 D_{\theta}}$.

\subsection{Coupling migration and switching dynamics}
 Combining our results in the previous sections, we can easily formulate a system of PDEs
  \begin{align}
     \partial_t \rho_0 (x,t) &= \nu E(\rho_0,\rho_1) \\
     \partial_t \rho_1 (x,t) &=  D_1 \laplacian{\rho_1}-\nu E(\rho_0,\rho_1)\\
      E(\rho_0, \rho_1) &= \bar{W}_{10}(\rho_0, \rho_1) \rho_1 - \bar{W}_{01}(\rho_0, \rho_1)\rho_0 
 \end{align} 
 where $E(\rho_0,\rho_1)$ is the phenotypic exchange term and $\nu$ is the corresponding timescale ratio of the switching and diffusion processes, i.e. $\nu=\frac{\tau_S}{\tau_D}$. To ensure the numerical consistency of the above system, we assume that resting cells diffuse in a very slow manner i.e. $D_{0} \ll D_{1}$, which results in the following reaction-diffusion system of equations
 \begin{align}
     \partial_t \rho_0(\vb{x}, t) &= D_{0}\laplacian{\rho_0} +\nu E(\rho_0, \rho_1),
     \label{simon1}
     \\
     \partial_t \rho_1(\vb{x}, t) &= D_{1} \laplacian{\rho_1} - \nu E(\rho_0, \rho_1).
     \label{simon2}
 \end{align}

\section{Spatio-temporal dynamics of the LEUP-driven migration/proliferation plasticity}

In this section, we study the mean-field approximation of the aforementioned stochastic plasticity dynamics for different regimes of the sensitivity $\beta$ and the interaction radius $\ell$.
In particular, we initially study the system dynamics, when cells have a very large interaction radius ($\ell\gg1 )$ and then a finite one.
Finally, we study  also the special case $\beta=0$, i.e. cells decide independently of their microenvironment. 
\subsection{Large interaction radius case: existence of a critical sensitivity}
Here, we focus on the very large interaction radius system ($\ell\gg1\implies V\gg 1$).
Although, this parameter regime is not biologically relevant, it is very instructive since it allows us to derive analytical estimates for our systems dynamics.
Here, we generalize the macroscopic system by adding proliferation dynamics (logistic growth).
Our phenotypic switch dynamics are recapitulated by setting the proliferation rate to zero, i.e. $r=0$. The full system reads
\begin{equation}
    \begin{split}
        & \frac{\partial \rho_{0}}{\partial t} =D_{0}\nabla^{2}{\rho_{0}}+\nu E_{0}\left(\rho_{0},\rho_{1}\right)+r\rho_{0}\left(1-\rho_{1}-\rho_{0}\right), \\
        & \frac{\partial \rho_{1}}{\partial t} =D_{1}\nabla^{2}{\rho_{1}}-\nu E_{0}\left(\rho_{0},\rho_{1}\right).
        \label{diff}
    \end{split}
\end{equation}
 In turn, we conduct a non-dimensionalization  of eqns.~(\ref{diff}) to identify the variables. Moreover, this helps us to gain a knowledge about the relationships between the different model parameters.
 By assuming that the system size is fixed at $L$ the non-dimensional quantities read 
\begin{equation}
\begin{split}
x^{*} &=\frac{x}{L}\implies\frac{\partial^{2}}{\partial x^{*2}}=L^{2}\frac{\partial^{2}}{\partial x^{2}},\\
t^{*} &=\frac{D_{0}t}{L^{2}}\implies\frac{\partial}{\partial t^{*}}=\frac{L^2}{D_0}\frac{\partial}{\partial t},\\
\gamma &= \frac{L^{2}\nu}{D_0},\\
D &= \frac{D_{1}}{D_{0}},\\
r^{'} &=\frac{r}{\nu}.
\end{split}
\end{equation} 
In the limit $V\gg 1$, the eqn.~($\ref{diff}$) can also be written as
\begin{equation}
    \begin{split}
        & \frac{\partial \rho_{0}}{\partial t^{*}} =\nabla^{*2}{\rho_{0}}+\gamma\left(\left(\rho_{1}-\rho_{0}\right)\left(\frac{1}{2}-\tilde{\beta}\frac{\left(\rho_{1}+\rho_{0}\right)}{\rho_{1}\rho_{0}}\right)+r^{'}\rho_{0}\left(1-\rho_{1}-\rho_{0}\right)\right),\\
        & \frac{\partial \rho_{1}}{\partial t^{*}} =D\nabla^{*2}{\rho_{1}}-\gamma\left(\left(\rho_{1}-\rho_{0}\right)\left(\frac{1}{2}-\tilde{\beta}\frac{\left(\rho_{1}+\rho_{0}\right)}{\rho_{1}\rho_{0}}\right)\right).
        \label{diffp}
    \end{split}
\end{equation}
To understand the behaviour of the system at long times, we conduct a fixed point analysis. Initially we assume a well-stirred system, i.e. no spatial interactions. Then, eqs.~(\ref{diffp}) can be written as coupled non-linear ODEs which have three fixed points
\begin{equation}
\begin{split}
&\left(\rho_{0}^{A*}, \rho_{1}^{A*}\right)=\left(\frac{1}{2}, \frac{1}{2}\right),\\ &\left(\rho_{0}^{B*}, \rho_{1}^{B*}\right)=\left(\frac{1}{2}\left(1+\sqrt{ 1-8\tilde{\beta}}\right),\frac{1}{2}\left(1-\sqrt{ 1-8\tilde{\beta}}\right)\right),\\
& \left(\rho_{0}^{C*}, \rho_{1}^{C*}\right)=\left(\frac{1}{2}\left(1-\sqrt{ 1-8\tilde{\beta}}\right),\frac{1}{2}\left(1+\sqrt{ 1-8\tilde{\beta}}\right)\right).
\end{split}
\label{eqbi}
  \end{equation}

The above imply a pitchfork bifurcation for the $\tilde{\beta}=\frac{\beta}{8V}$ parameter, i.e. there exists a critical value $\tilde{\beta}_{c}$ that introduces the bistable state.
From eqs.(\ref{eqbi}), we can easily deduce that the critical sensitivity value 
\begin{equation}
    1-8\tilde{\beta_{c}}=0\Longleftrightarrow\beta_{c}=V.
\end{equation}
This is an acceptable approximation even for the finite interaction radius system (see next section). 
In the following, we analyze if the system is able to produce spatial patterns.
Applying linear stability analysis for the spatially resolved system eqs.~(\ref{diffp}), we can deduce that no pattern formation is possible (for details check next section).
Any perturbations to the homogeneous state lead always to a spatially homogeneous steady state. 
When $r=0$ this result is consistent with our findings in discrete IBM simulations where very large interaction radii do not confer any clustering, as shown in Fig. \ref{radial2}. 

\subsection{The finite interaction radius case: emergence of a bistable switch between "fluid" and "solid" tissue phases and pattern formation}
\begin{figure}[!htb]
  \centering
 \begin{subfigure}[b]{0.45\linewidth}
\includegraphics[width=\linewidth]{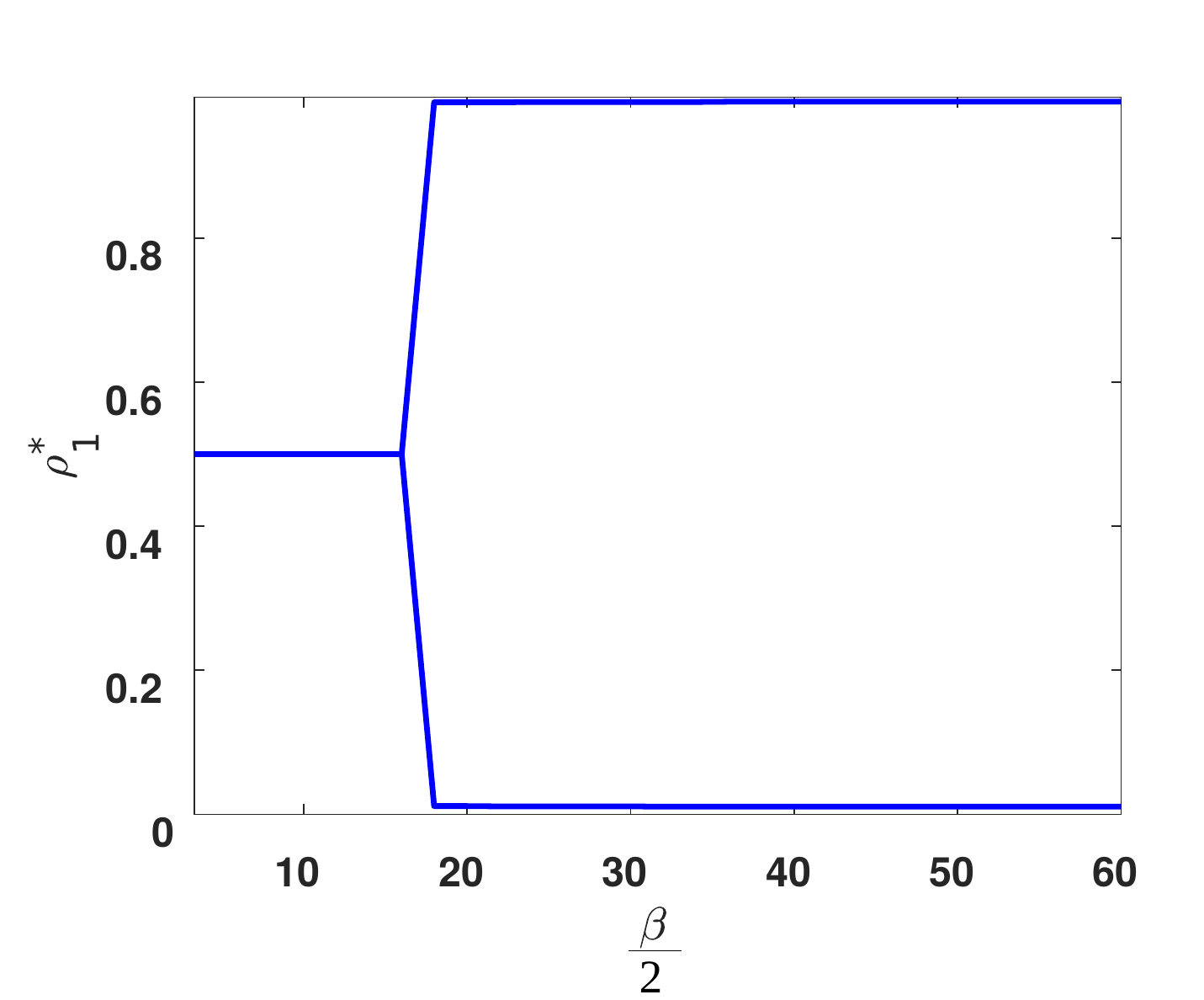}
\caption{}\label{bifur1}
\end{subfigure}
~
\begin{subfigure}[b]{0.45\linewidth}
\includegraphics[width=\linewidth]{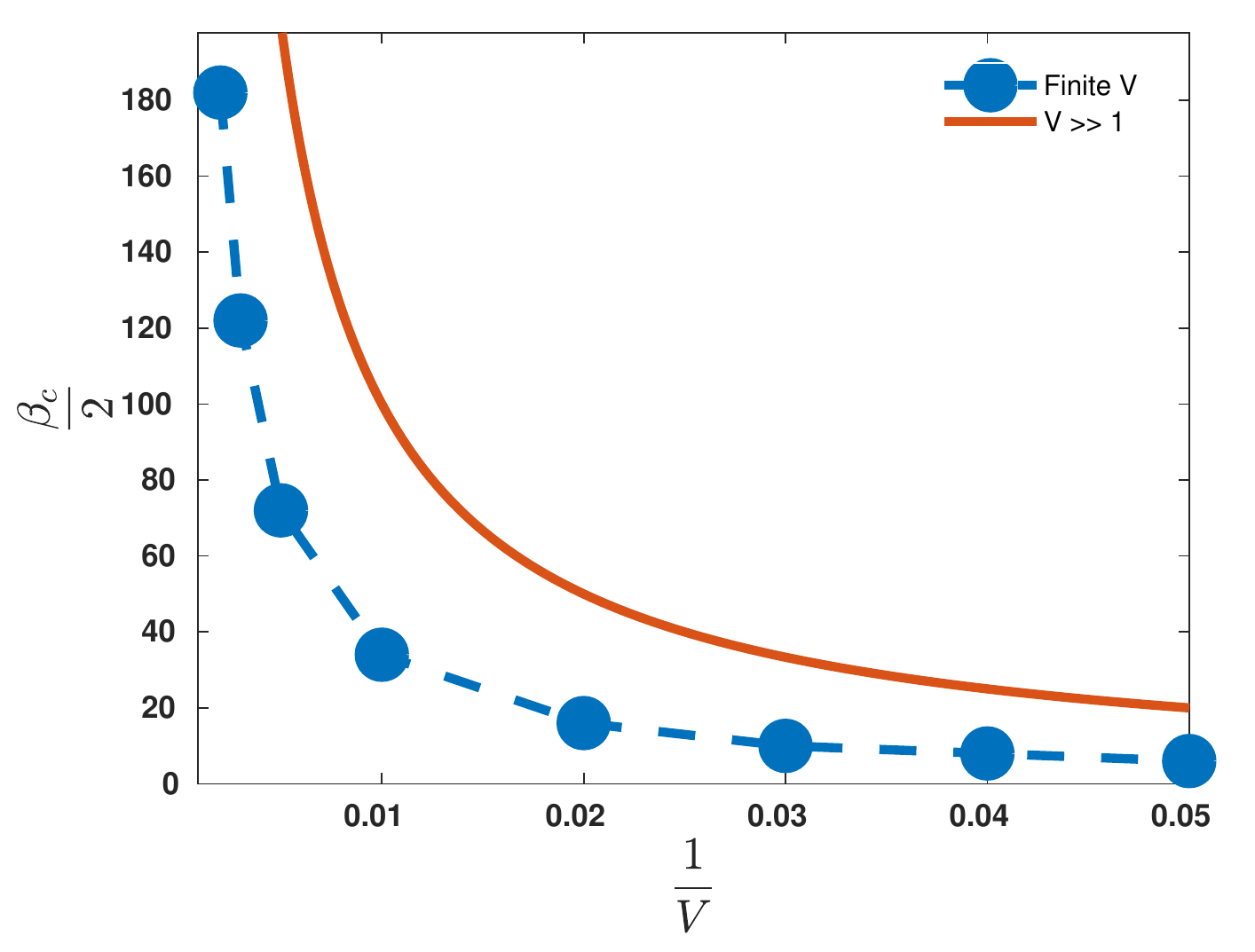}
\caption{}\label{crit}
\end{subfigure}
 \caption{(\textbf{a}) Bifurcation diagram for moving cells in the finite interaction radius limit with respect to $\beta$, where  proliferation rate $r$ is fixed to 0 and $V^{-1}$ at 0.02.
 The bifurcation diagram is symmetric for resting cells i.e. $\rho_{0}^{*}$.
 (\textbf{b}) The critical point value of the bifurcation diagram (where the solution splits in two branches) is plotted against the inverse volume  $\frac{1}{V}$.}
 \label{fig:bifur}
\end{figure}

Now we turn to the full system for intermediate interaction radius also (assuming proliferation). This implies a finite interaction volume $V$ and  for analytical feasibility we are interested in the Gaussian approximation of the switching probabilities and their corresponding mean-field terms. The full system of PDEs assuming also proliferation reads 
\begin{equation}
\begin{split}
    \frac{\partial \rho_{0}}{\partial t^{*}} &= \nabla^{*2}{\rho_{0}}+\gamma\left( \frac{\rho_{1}-\rho_{0}\left(\frac{\rho_{0}\left(\rho_{1}-\frac{1}{V}\right)}{\rho_{1}\left(\rho_{0}-\frac{1}{V}\right)}\right)^{\frac{\beta}{2}}}{1+\left(\frac{\rho_{0}\left(\rho_{1}-\frac{1}{V}\right)}{\rho_{1}\left(\rho_{0}-\frac{1}{V}\right)}\right)^{\frac{\beta}{2}}}+r^{'}\rho_{0}\left(1-\rho_{1}-\rho_{0}\right)\right),\\
        \frac{\partial \rho_{1}}{\partial t^{*}} &= D\nabla^{*2}{\rho_{1}}-\gamma\frac{\rho_{1}-\rho_{0}\left(\frac{\rho_{0}\left(\rho_{1}-\frac{1}{V}\right)}{\rho_{1}\left(\rho_{0}-\frac{1}{V}\right)}\right)^{\frac{\beta}{2}}}{1+\left(\frac{\rho_{0}\left(\rho_{1}-\frac{1}{V}\right)}{\rho_{1}\left(\rho_{0}-\frac{1}{V}\right)}\right)^{\frac{\beta}{2}}}.
        \label{meanfin}
\end{split}
\end{equation}
Since closed expression of steady states are not analytically feasible, we obtain the bifurcation diagram numerically, see Fig~(\ref{bifur1}). 
 We observe the existence of a supercritical pitchfork bifurcation and the existence of a critical $\beta_c$.
 For $\beta\geq\beta_c$, the systems depart from balanced state ($\rho_0=\rho_1=\frac{1}{2})$ to the coexistence of a \enquote{fluid} (most cells migrate) and a \enquote{solid} (most cells are resting) phase.
 The switch is controlled by the perturbation on the ratio of migratory and resting cells. Interestingly, we can compare the analytic estimate of $\beta_c = V$, from the $V\gg1$ case, with the one calculated for the finite $V$ system.
 Fig.~(\ref{crit}) shows that the infinite system approximation provides an upper bound for $\beta_c$, which is not too far from the real value of finite systems. 
 
The existence of a critical $\beta_c$ with respect to the two phases is evident in the IBM simulations as well. In particular if we quantify the ratio of stationary over motile cells, we observe a similar behavior of the critical sensitivity for increasing $V$, since it decreases (see Fig.\ref{critibm} in SI ). However, we cannot conduct a strict quantitative comparison since our bifurcation analysis does not involve any diffusion as opposed to the IBM simulation.
 
\begin{figure}[htb]
    \centering 
\begin{subfigure}{0.3\textwidth}
  \includegraphics[width=\linewidth]{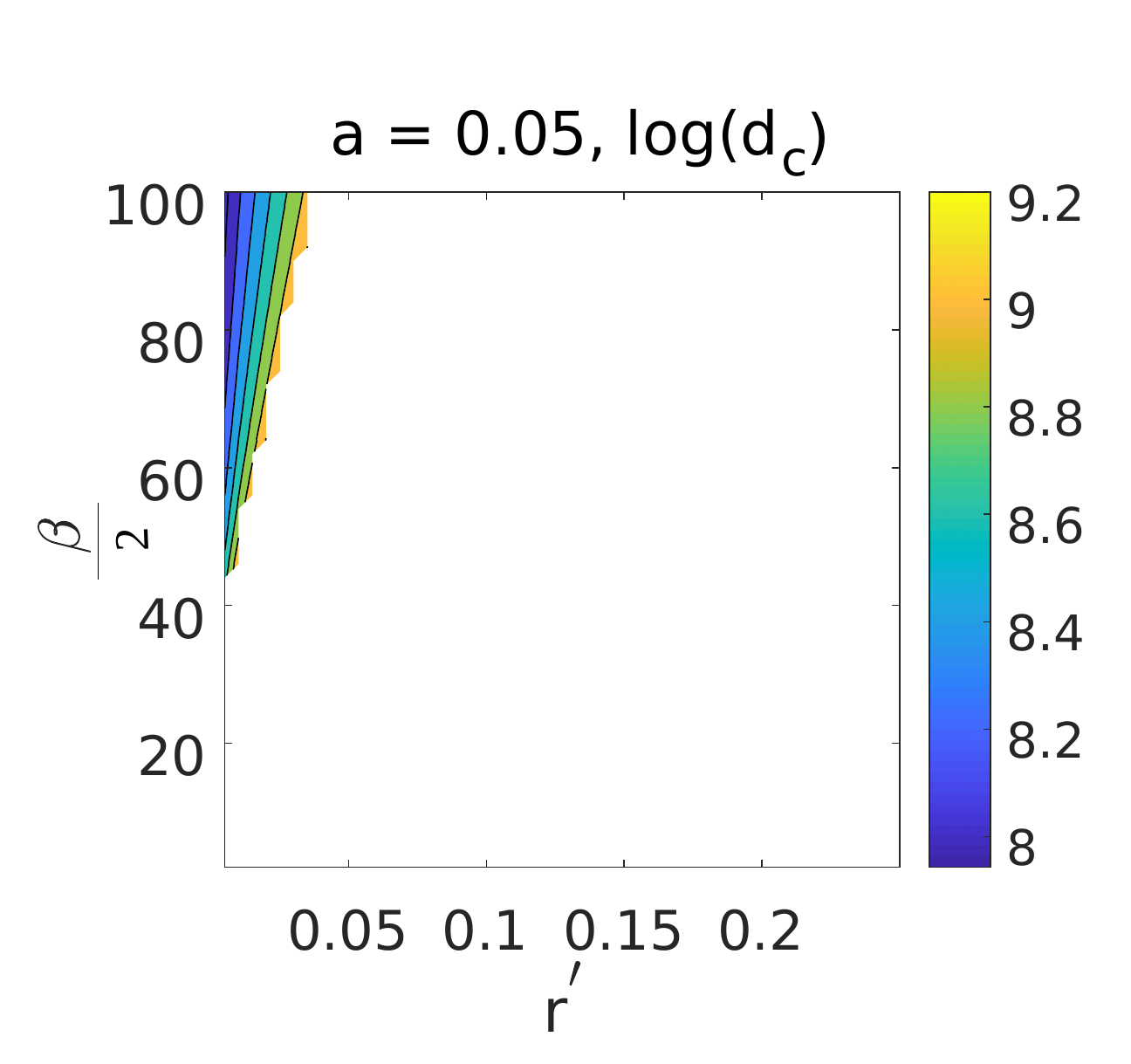}
  \label{fig1}
\end{subfigure}\hfil 
\begin{subfigure}{0.3\textwidth}
  \includegraphics[width=\linewidth]{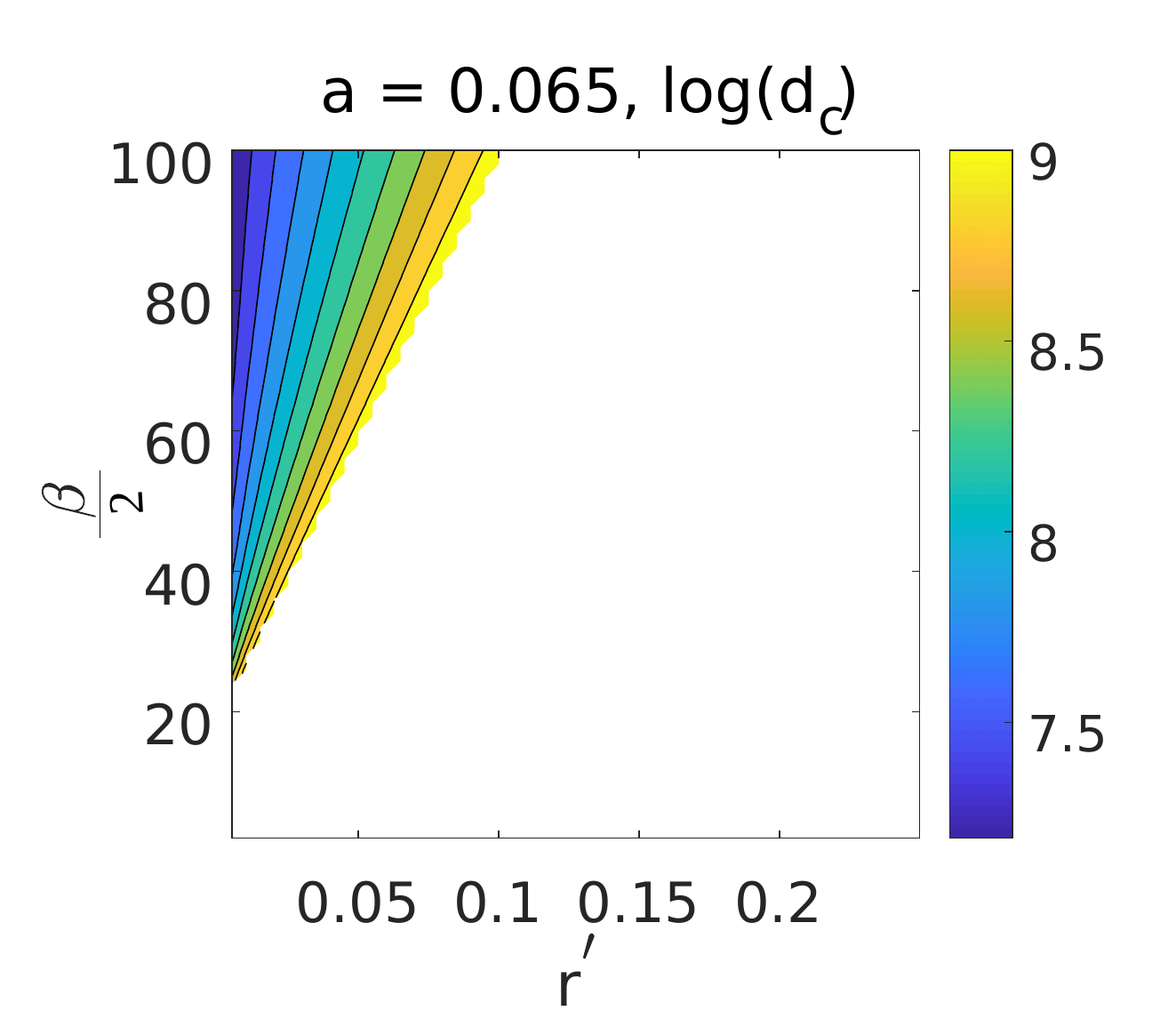}
  \label{fig2}
\end{subfigure}\hfil 
\begin{subfigure}{0.3\textwidth}
  \includegraphics[width=\linewidth]{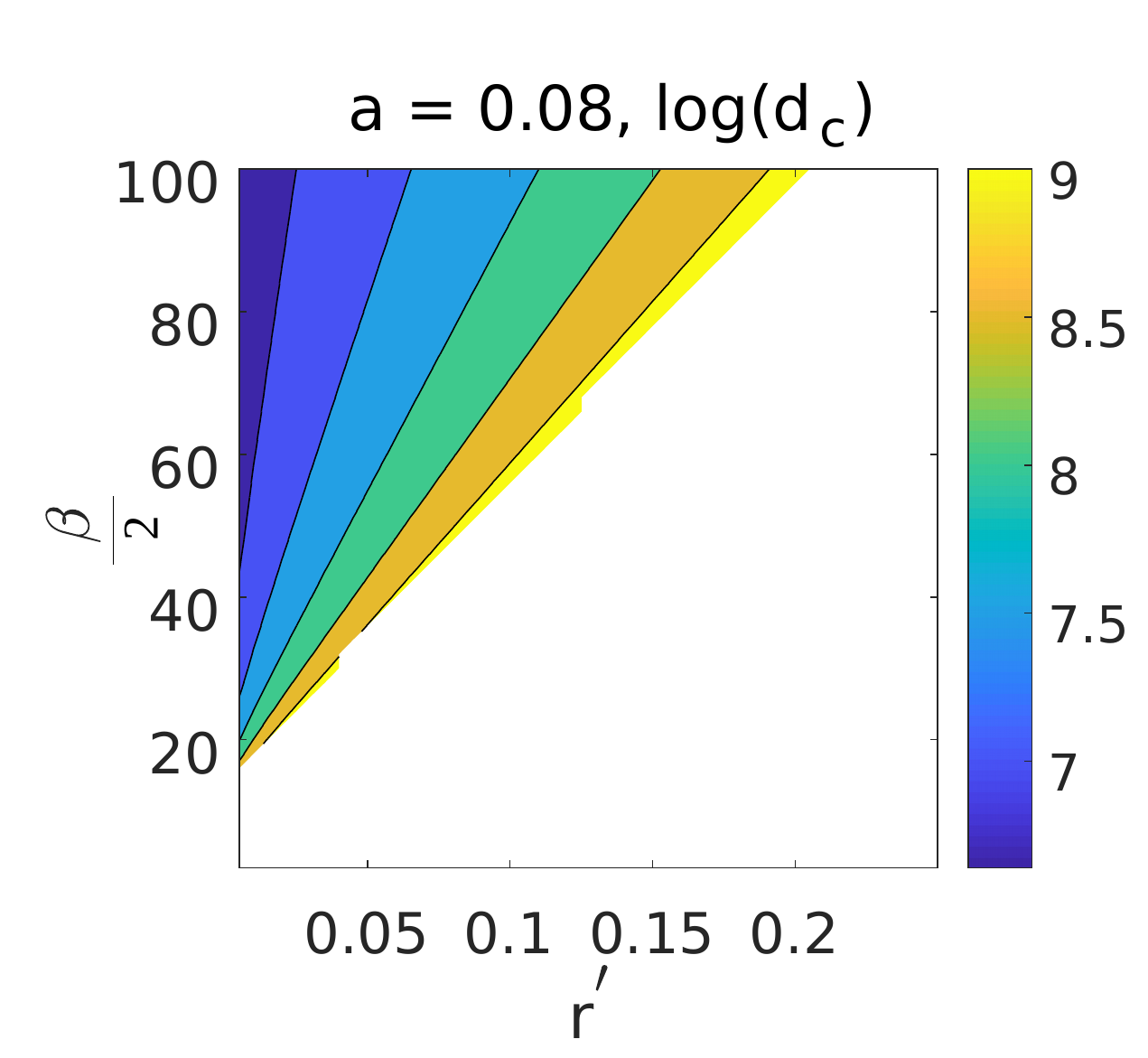}
  \label{fig3}
\end{subfigure}

\medskip
\begin{subfigure}{0.3\textwidth}
  \includegraphics[width=\linewidth]{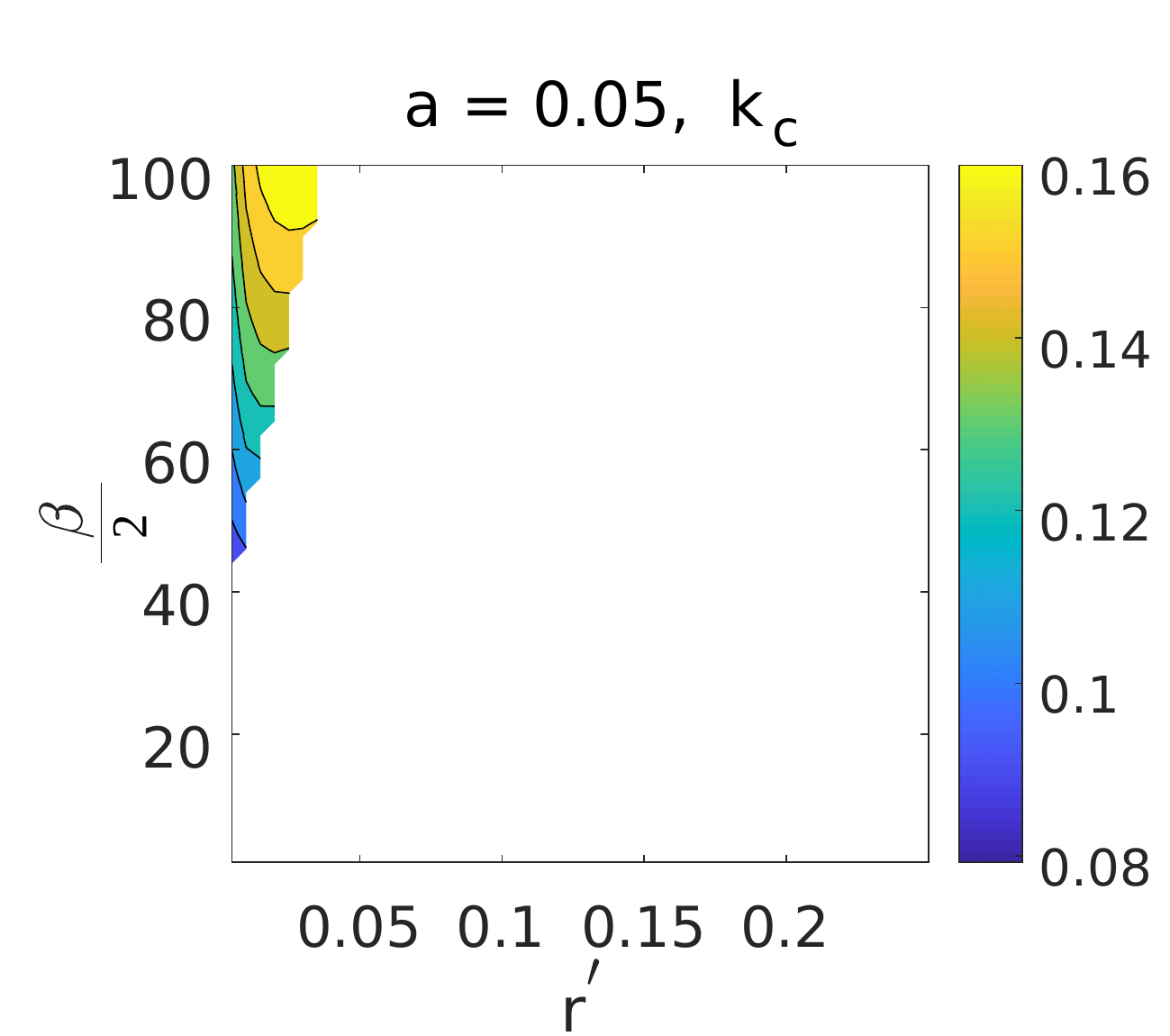}
  \label{fig4}
\end{subfigure}\hfil
\begin{subfigure}{0.3\textwidth}
  \includegraphics[width=\linewidth]{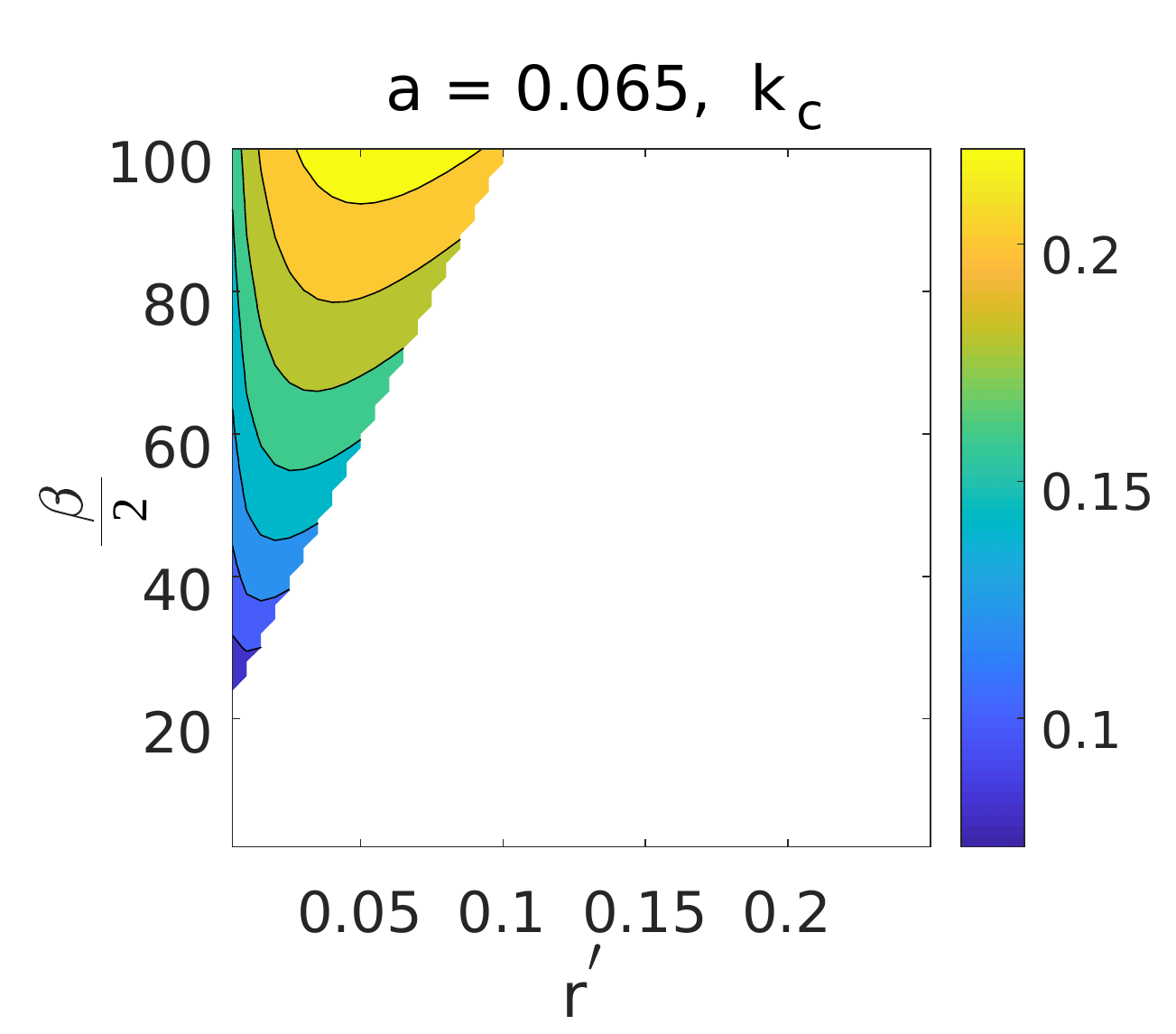}
  \label{fig5}
\end{subfigure}\hfil 
\begin{subfigure}{0.3\textwidth}
  \includegraphics[width=\linewidth]{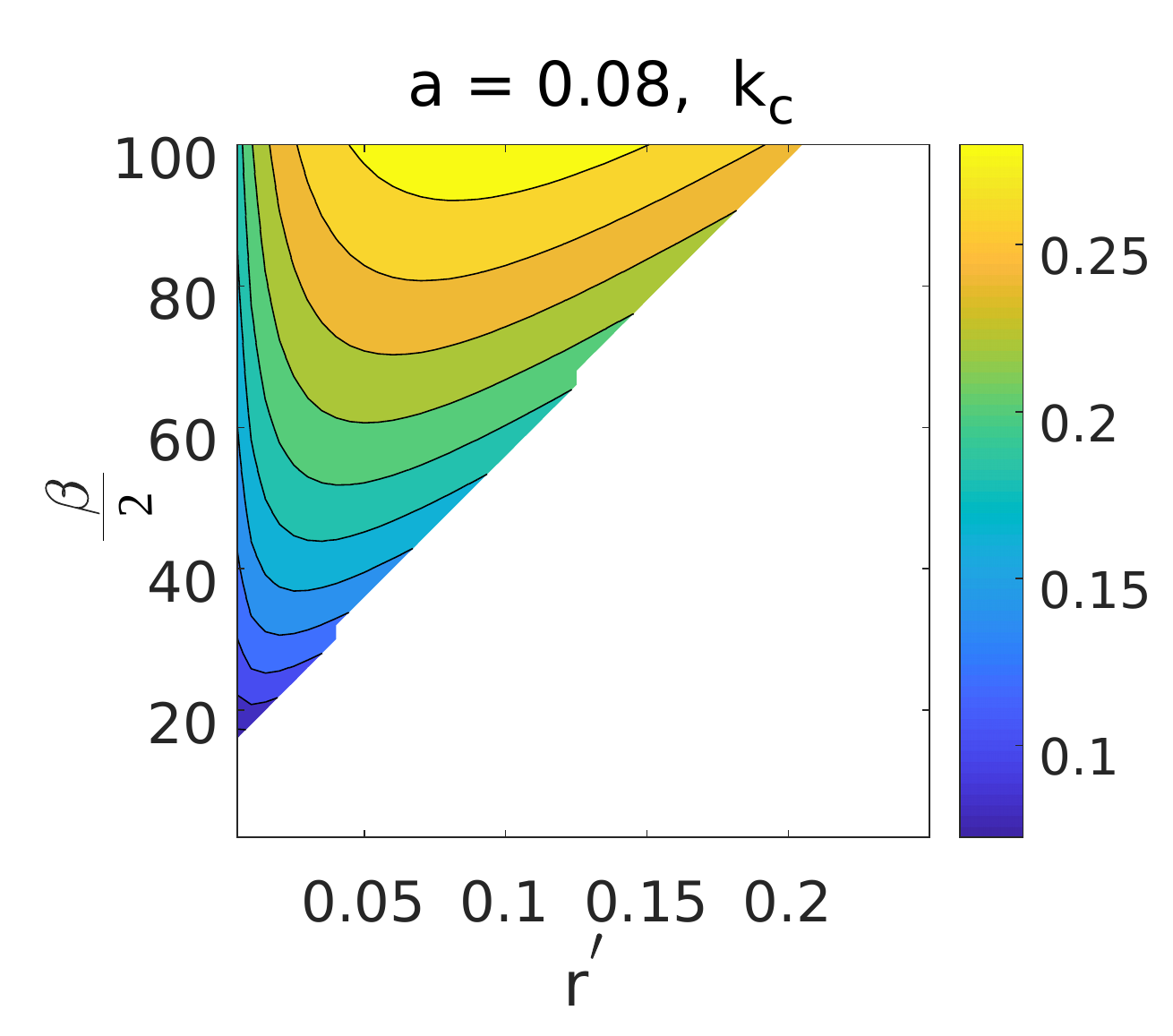}
  \label{fig6}
\end{subfigure}\hfil 
\caption{Turing space diagram of sensitivity and proliferation rate, i.e. $\frac{\beta}{2}$ vs. $r^{'}$, where different values of $\log{(d_{c})}$ and $k_{c}$ have been plotted. Here we denote  $a=\frac{1}{V}$.} \label{fig:1}
\end{figure}
\newpage
In turn, we apply linear stability analysis to identify parameter regimes that promote diffusion driven pattern formation or Turing instability. To analyze the Turing instability \cite{Murray2003} we have to find the system's steady state,  (i.e. when diffusion is not present in the systems of eqn.(\ref{diff})).
 It has been shown that $d=\frac{D_1}{D_0}\gg1$ is a necessary condition for the emergence of spatially heterogeneous solution i.e. patterns. 
Now, we can write the system of PDEs (i.e eqn.(\ref{diff})) in a generalized matrix form
\begin{equation}
   \begin{split}
\frac{\partial \boldsymbol{\rho}}{\partial t} &= \boldsymbol{D}\nabla^{2}\boldsymbol{\rho}+\gamma\boldsymbol{R\rho},\hspace{0.25cm} \boldsymbol{D}= \begin{pmatrix}
1 & 0 \\
0 & d 
\end{pmatrix},
 \hspace{0.25cm}\\
\boldsymbol{R} &= \begin{pmatrix}
\frac{\partial E_{0}\left(\rho_{0},\rho_{1}\right)}{\partial \rho_0}+r^{'}\left(1-\rho_{1}-\rho_{0}\right)-r^{'}\rho_{0} & \frac{\partial E_{0}\left(\rho_{0},\rho_{1}\right)}{\partial \rho_1}-r^{'}\rho_{0} \\
-\frac{\partial E_{0}\left(\rho_{0},\rho_{1}\right)}{\partial \rho_0} & -\frac{\partial E_{0}\left(\rho_{0},\rho_{1}\right)}{\partial \rho_1}
\end{pmatrix}
_{\left(\rho_0^{*},\rho_1^{*}\right)},\\
& = \begin{pmatrix}
f_{u} & f_{v}\\
g_{u} & g_{v}
\end{pmatrix}_{\left(\rho_0^{*},\rho_1^{*}\right)},
 \end{split}
 \label{part}
\end{equation}
where $\boldsymbol{\rho}$ is defined as $\left(\rho_{0}-\rho_0^{*},\rho_{1}-\rho_{1}^{*}\right)^{T}$ and $\boldsymbol{R}$ is the Jacobian at $\rho^{*}=(\rho_{0}^{*},\rho_1^{*})$. Using the Turing conditions of instability \cite{Murray2003}, we found patterns when $N$ is finite in zero-flux boundary conditions. We have checked all the Turing instability conditions:
\begin{equation}
    \begin{split}
        &f_{u}+g_{v}<0,\hspace{0.8cm}f_{u}g_{v}-f_{v}g_{u}>0,\\
        &D f_{u}+g_{v}>0,\hspace{0.8cm}(D f_{u}+g_{v})^{2}-4D(f_{u}g_{v}-f_{v}g_{u})>0.
    \end{split}
    \label{condtu}
\end{equation}
Interestingly, only when the density of resting cells is larger than that of the moving cells, patterns are formed under Turing instability conditions. In order to investigate the system's potential to exhibit pattern formation, we check the range of validity of the Turing instability criteria (\ref{condtu}).
Diffusion-driven instability conditions are satisfied for a large portion of the parameter space.
In turn for these parameters, we calculate the critical diffusion coefficient $d_{c}$. For values $d>d_{c}$ we are able to observe patterns,
The existence of $d_{c}$ is associated with the existence of a critical wavenumber $k_{c}$ \cite{Murray2003}.
\begin{equation}
\begin{split}
    &d_c=\frac{-2(2f_{v}g_{u}-f_{u}g_{v})\pm\sqrt{\left(2(2f_{v}g_{u}-f_{u}g_{v})\right)^{2}-4f_{u}^{2}g_{v}^{2}}}{2f_{u}^{2}},\\
    &k_c=\gamma\left[\frac{Det(A)}{d_c}\right]^{\frac{1}{2}}.
    \end{split}
\end{equation}
In Fig.~(\ref{fig:1}) we identify the parameter regime that allows us to observe patterns and then corresponding $d_{c}$ and $k_{c}$.

\begin{figure}
    \centering
    \includegraphics[width=0.9\textwidth]{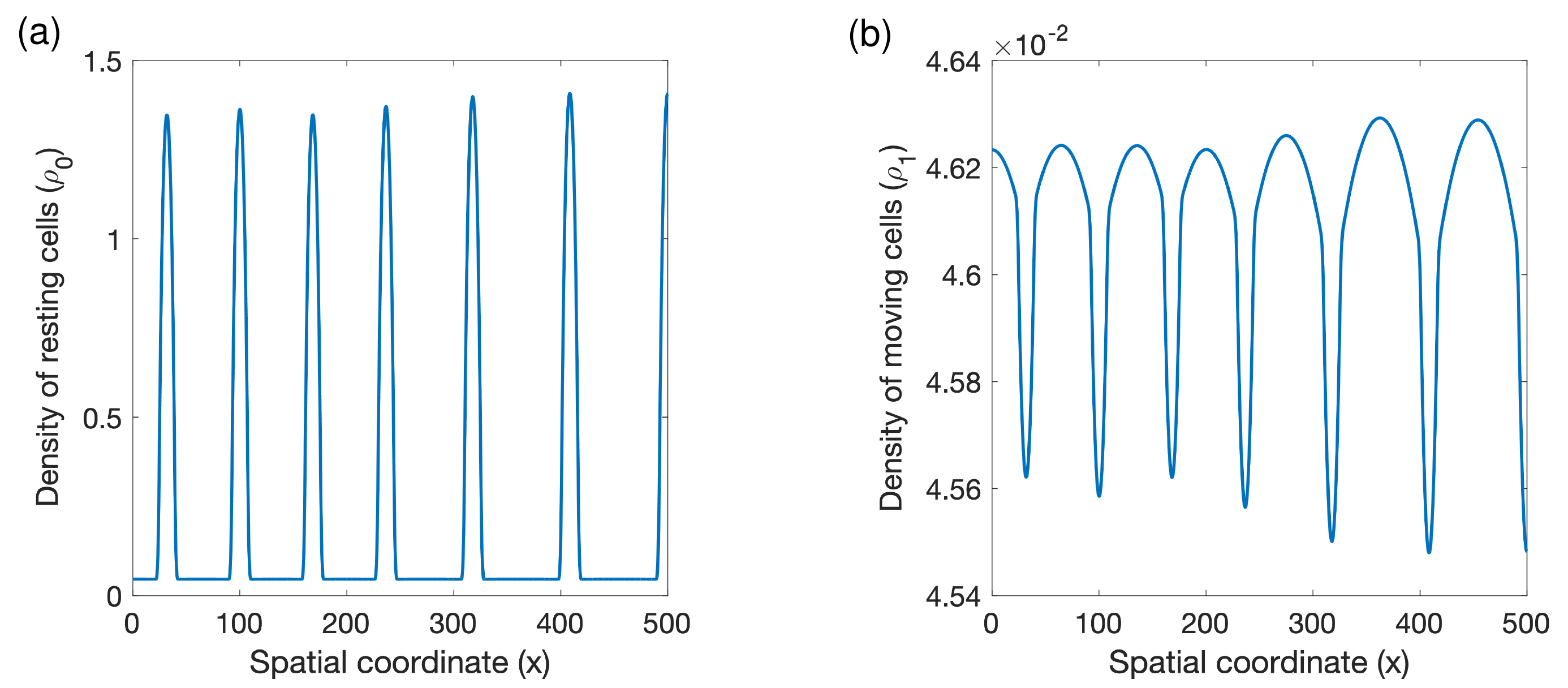}
    \caption{Densities of proliferating (\textbf{a}) and migrating (\textbf{b}) cells of the 1D mean-field simulation after 50000 time steps.}
    \label{fig:RhoTfin_1D}
\end{figure}

\begin{figure}
    \centering
    \includegraphics[width=0.9\textwidth]{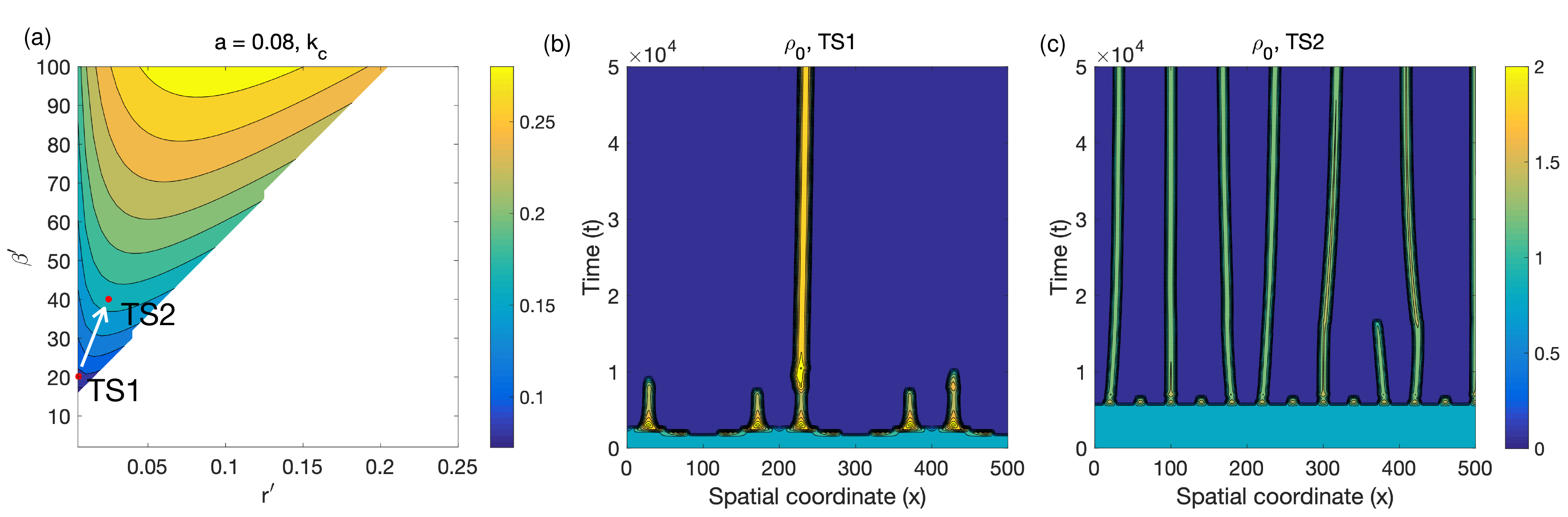}
    \caption{One-dimensional simulation of the mean-field approximation for finite interaction radius system. In (\textbf{b-c}) densities of proliferating cells are plotted over time corresponding to two points (TS1 and TS2) inside the Turing space shown in (\textbf{a}). Here $\beta^{'}=\frac{\beta}{2}$ and $a$ is defined as the inverse volume $\frac{1}{V}$.} 
    \label{fig:rho0_1D}
\end{figure}

By fixing the initial conditions to a cosine wave, we observe in one dimension the existence of regular spikes as shown in Fig~(\ref{fig:RhoTfin_1D}). Please note that the exact pattern is sensitive to the initial conditions.
In turn, we simulate our system in 2D and for the same parameters and we observe patterning in the form of dots.
Finally, we investigate if the type of patterns changes for variations in parameters $\beta$ and $r^{'}$ (see Fig.~(\ref{fig:rho0_1D})).
If $d>d_{c}$ then we need a large domain or size of the system to observe the patterns.
In the 2D case, we have observed the discoidal patterns of resting phenotype which resemble the 1D case.
So, the radius of the circles of the patterns are increased if we fix the domain size.
Moreover, we observe that in both dimensions (i.e. 1D and 2D) the critical spatial frequency increases with decreasing $r^{'}$. 
%
%
%
\begin{figure}
    \centering
    \includegraphics[scale=0.3]{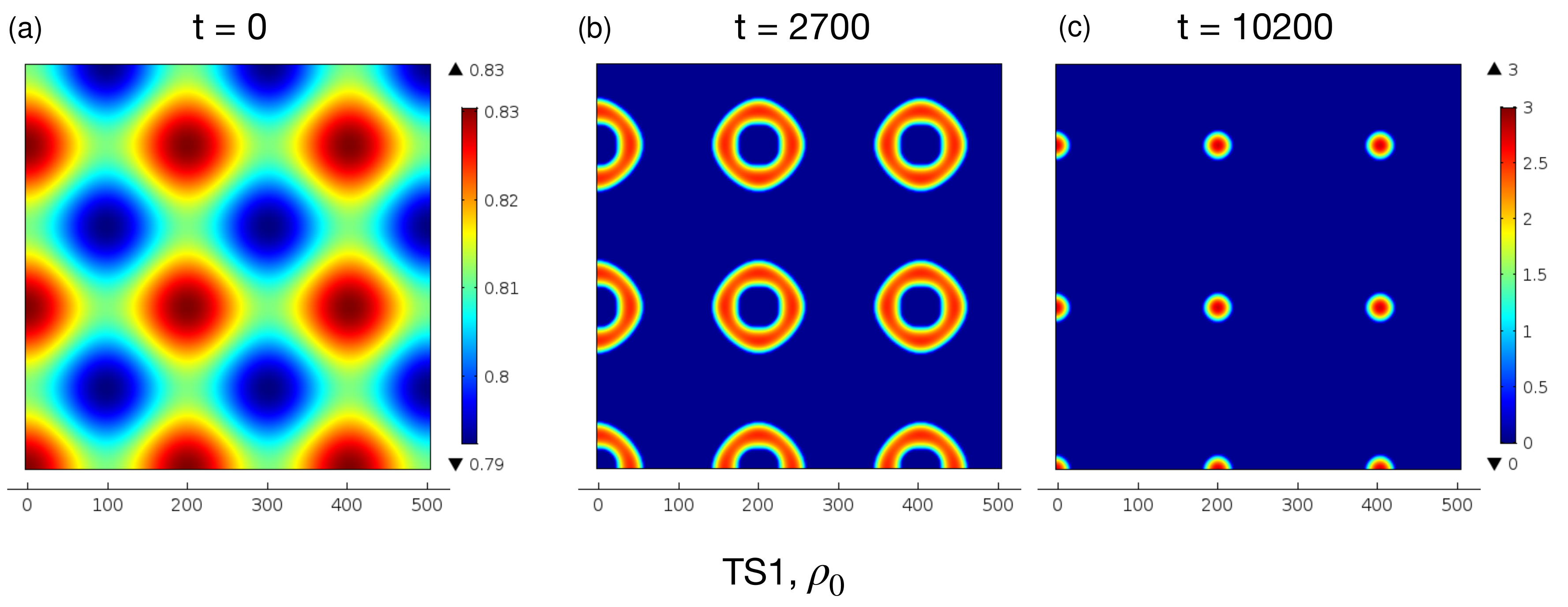}
    \caption{Two-dimensional simulation of the mean-field approximation for finite interaction radius system. The patterns are obtained for parameters that correspond to the first point (TS1)  in the Turing space, as displayed in Fig.(\ref{fig:rho0_1D})}
    \label{fig:my_label}
\end{figure}
In comparison with the discrete IBM simulations, we have to state that the simulation clusters  can be identified by the discoidal mean-field patterns. Under this statement, we find that:
\begin{enumerate}
    \item There exists a critical $\beta$ that allows for the emergence of patterns as in the IBM simulations.
        \item For very low and very high interaction radius $\ell$, we observe no patterns, which is consistent with our IBM results (see Fig.\ref{radial2}).
        
\end{enumerate}
\subsection{Microenvironment independent phenotypic switching leads to uncontrolled growth dynamics}
 Finally we investigate the case $\beta$ = 0 where cells do not sense their microenvironment and so we end up with the following system,
 \begin{equation}
\begin{split}
    & \frac{\partial \rho_{0}}{\partial t^{*}} =\nabla^{*2}{\rho_{0}}+\gamma\left(\frac{\left(\rho_{1}-\rho_{0}\right)}{2}+r^{'}\rho_{0}\left(1-\rho_{1}-\rho_{0}\right)\right),\\
        & \frac{\partial \rho_{1}}{\partial t^{*}} =D\nabla^{*2}{\rho_{1}}-\gamma\frac{\left(\rho_{1}-\rho_{0}\right)}{2}.
        \label{mean2}
\end{split}
\end{equation}
We can do a fixed point analysis similar to equation (\ref{diffp}).
Assuming no spatial dynamics, we can find two fixed points i.e.
\begin{equation}
\begin{split}
    \left(\rho_{0}^{P*}, \rho_{1}^{P*}\right) &=\left(\frac{1}{2}, \frac{1}{2}\right), \\
    \left(\rho_{0}^{Q*}, \rho_{1}^{Q*}\right) &=(0,0).
    \end{split}
\end{equation}
For further details see Fig.~\ref{phase} in S.I.
We can clearly show that the  fixed point (0,0) is unstable and $\left(\frac{1}{2},\frac{1}{2}\right)$ is a saddle point.
We can also write the coupled PDE equations in a single PDE as well, if we consider $\rho_{0}=\rho_{1}=\tilde{\rho}$. 
Then we obtain
\begin{equation}
    \frac{\partial \tilde{\rho} }{\partial t^{*}} =\frac{(1+D)}{2}\nabla^{*2}{\tilde{\rho}}
+\frac{r}{2}\tilde{\rho}\left(1-2\tilde{\rho}\right).
\label{mean3}
\end{equation}
Now we can clearly see that the eqn.(\ref{mean3}) is similar to the Fisher-Kolmogorov equation \cite{Murray2003}.
It is known that the Fisher-Kolmogorov equation does not exhibit any pattern formation instabilities. The latter observation is consistent with our discrete IBM simulations, where no clustering is observed.

\newpage
\section{Discussion}
Recently, the Least microEnvironmental Uncertainty Principle (LEUP) has been proposed
as an organization principle for  cell decision-making in multicellular systems. In this paper, we apply this principle to shed light on the effects of phenotypic plasticity for tissue dynamics with a mathematical model. We focus on two types of plasticity: a go or rest, and a go or grow phenotypic dichotomy, which play key roles in important processes in biological development and pathological situations as cancer invasion. We  assume that in any given spatial mesoscopic sample the presence of cell phenotypes follows a binomial distribution. Using this assumption, we are able to calculate the microenvironmental entropy and the LEUP-driven probability distribution of each phenotype. On the basis of this distribution, we defined an appropriate microscopic stochastic model for the spatio-temporal dynamics of both phenotypes, and in turn derived the corresponding mean-field description resulting in a system of coupled reaction-diffusion equations. The main results of our study are: (i) in the case of go or rest plasticity, there exists a supercritical pitchfork bifurcation that defines a switch between a "fluid" and a "solid" phase, and (ii) in the case of go or grow plasticity, for the "solid" phase, we can derive conditions for the emergence of Turing patterns. Interestingly, the existence of Turing patterns requires a critical LEUP sensitivity to the microenvironment and a minimum interaction range. 

Our model assumes binary transitions between two discrete phenotypes, which is an extreme simplification of biological reality. Certainly, one could include a continuous spectrum of motile/proliferative phenotypes that is expected to imply even richer spatio-temporal dynamics. Indeed, assuming a continuous state (velocity) space would potentially lead to further interesting bifurcations, such as the metastable EMT state as found in  \cite{Jollyrsc, BoaretoE402}, or complex spatiotemporal patterns as indicated in \cite{Barua404889}.
But the very abstracted case chosen here is still instructive to indicate how entropy-driven phenotypic decisions can lead to particular types of spatio-temporal pattern formation. 

Interestingly, our LEUP-driven IBM is an extension of a Vicsek-type  model \cite{vicsek1995}, formulated in the context of self-propelled particles \cite{grossmann2013}.
Our model exhibits a novel collective behavior when compared to the past published results from Vicsek-type of models.
In particular, to our knowledge it is the first time to produce with such models Turing patterns, i.e. dynamics clusters of non-motile cells of specific characteristic wavelength.
Typically, in Viscek-type models we may observe moving clusters of swirling cells (e.g. the milling Viscek model) but never static ones. 

At this point, let us focus on the biological assumptions and implications of our study. The molecular regulatory mechanisms involved in EMT or GoG remain largely unknown, where the latter can be viewed as an EMT with proliferation constrained to the epithelial/resting phase.
Here, we assume that the phenotypic regulation of both mechanisms is based on the minimization of microenvironmental entropy in physiological tissues, which allows us to predict the multicellular spatiotemporal dynamics.
This assumption is supported by the fact that healthy physiological processes, biological development or processes like wound healing, where EMT/MET or GoG are present, typically lead to an ordered (low entropy) tissue from a disordered initial condition.
On the other hand, deregulation of EMT and GoG have been already identified as pivotal elements in invading cancers \cite{banyard,Hatzikirou2010a}, where genetic and phenotypic heterogeneity, characterized by high entropy, is a key characteristic.
Assuming a LEUP-driven migration/proliferation phenotypic regulation allows us to understand how cells control multicellular dynamics in terms of growth and patterning.

\textit{Implications in multicellular growth control:} The central finding of our study is associated with the bifurcation diagram of the LEUP sensitivity parameter $\beta$ in Fig.~\ref{bifur1}. As stated before, the parameter $\beta$ quantifies how prone cells are in sensing and responding to their microenvironmental stimuli. When $\beta=0$ cells migrate and proliferate independent of their microenvironmental cues, which corresponds to one of the cancer hallmarks \cite{Hanahan2000}. In this case, the systems grows uncontrollably, resembling a cancerous tissue \cite{Hanahan2000}. The resulting Fisher-Kolmogorov macroscopic behavior has been prototypically used to model invading tumours \cite{dahlman}. On the other hand, by adding a death process for any motility state in the GoG model, we can recapitulate the Allee effect (bistability between extinction and growth) as found in Boettger et al \cite{Bottger2015}. A cell population at the “fluid” state will go extinct (motile cells do not proliferate but still have a probability to die,) where as systems in the “solid” state will always grow until carrying capacity. Therefore, in the bistability regime cell sensing properties lead to multicellular growth control.

\textit{Implications in multicellular pattern control:} Increased cell sensing $\beta$ represents the physiological tissue dynamics, since it allows the system to control its behavior. By tuning $\beta$ and the ratio of motile/resting cells, the system exhibits a bistable behavior between a "fluid"/mesenchymal-type and a “solid”/epithelial-type tissue phase. This kind of tissue level switch is of utmost importance in physiological processes such as wound healing or embryogenesis \cite{Barriere2015}.
After a tissue injury, the healing is characterized by a "fluid"  diffusive expansion of fibroblast cells, that adopt a migratory phenotype via EMT \cite{Yan}.
After covering the wound, the "solid" phase emerges as cells stop the migration program and proliferate to finalize tissue repair. In the abscence of proliferation, the bistable switch from the "solid" to the "fluid" state could potentially explain the jamming phase transition observed in epithelial colonies, under EGF modulation and Rab5a knock-out \cite{Malinverno2017}. Typically, the "solid" multicellular phase is prone to the emergence of pattern formation, which frequently occurs in physiological epithelial tissues \cite{Yan, Heller2015}. When EMT  is combined with a Notch-Delta cell-cell communication then epithelial/immotile cell clusters emerge \cite{BoaretoE402}, as observed in our IBM and mean-field simulations.  Adding proliferation in the GoG model, the type and the size of such emerging patterns require a tight regulation of microenvironmental sensing $\beta$ and proliferation rate $r$, as indicated by Fig.~\ref{turing0.01wc}, since the critical wavelength $k_c$ depends on the ratio $\frac{\beta}{r^{'}}$. Such Turing patterns are in agreement with previous GoG studies \cite{Pham2011}, where Turing patterns where emerging. 

In conclusion, our study shows how individual LEUP-driven cell decisions the dynamics at the tissue level and how knowledge of collective cell decision-making can be used to control of growth and pattern.

\subsection*{Acknowledgment} AB thanks the International Graduate School of HZI, Braunschweig. HH, MMH and PM gratefully acknowledge the funding support of the Helmholtz Association of German Research Centers—Initiative and Networking Fund for the project on Reduced Complexity Models (ZT-I-0010). HH and PM acknowledge the funding support of MicMode-I2T (01ZX1710B) by the Federal Ministry of Education and Research (BMBF). HH is supported by  MulticellML (01ZX1707C) of the Federal Ministry of Education and Research (BMBF) and the Volkswagenstiftung for the its support within the "Life?" programm (96732). SS acknowledges financial support co-financed by the European Social Fund (ESF) and tax funds in accordance with the budget adopted by the members of the Saxon State Parliament. Part of the current work was inspired and initiated when NK was visiting the  Helmholtz Centre for Infection Research and he would like to express his gratitude for the warm hospitality of the institute. NK would also like to acknowledge financial support from the Faculty of Science and Engineering of University of Chester. The authors thank the Centre for Information Services and High Performance Computing (ZIH) at TU Dresden for providing an excellent infrastructure.

\printbibliography

@article{Pham2011,
author = {Pham, Kara and Chauviere, Arnaud and Hatzikirou, Haralambos and Li, Xiangrong and Byrne, Helen M. and Cristini, Vittorio and Lowengrub, John},
doi = {10.1080/17513758.2011.590610},
issn = {1751-3758},
journal = {J. Bio. Dyn.},
month = {06},
number = {June},
pages = {1--18},
publisher = {Taylor {\&} Francis},
title = {{Density-dependent quiescence in glioma invasion: instability in a simple reaction–diffusion model for the migration/proliferation dichotomy}},
url = {http://www.tandfonline.com/doi/abs/10.1080/17513758.2011.590610},
year = {2011}
}

@article{Heller2015,
author = {Heller, Evan and Fuchs, Elaine},
doi = {10.1083/jcb.201506106},
issn = {15408140},
journal = {J. Cell Bio.},
number = {2},
pages = {219--231},
pmid = {26504164},
title = {{Tissue patterning and cellular mechanics}},
volume = {211},
year = {2015}
}

@article{Bottger2015,
author = {B{\"{o}}ttger, Katrin and Hatzikirou, Haralambos and Voss-B{\"{o}}hme, Anja and Cavalcanti-Adam, Elisabetta Ada and Herrero, Miguel A. and Deutsch, Andreas},
doi = {10.1371/journal.pcbi.1004366},
editor = {Alber, Mark S},
issn = {1553-7358},
journal = {PLOS Comp. Bio.},
month = {09},
number = {9},
pages = {e1004366},
pmid = {26335202},
title = {{An Emerging Allee Effect Is Critical for Tumor Initiation and Persistence}},
url = {http://dx.plos.org/10.1371/journal.pcbi.1004366},
volume = {11},
year = {2015}
}

@article{vicsek1995,
  title={Novel type of phase transition in a system of self-driven particles},
  author={Vicsek, Tam{\'a}s and Czir{\'o}k, Andr{\'a}s and Ben-Jacob, Eshel and Cohen, Inon and Shochet, Ofer},
  journal={Phys. rev. lett.},
  volume={75},
  number={6},
  pages={1226},
  year={1995},
  publisher={APS}
  }

@article{grossmann2013,
  title={Self-propelled particles with selective attraction--repulsion interaction: from microscopic dynamics to coarse-grained theories},
  author={Grossmann, Robert and Schimansky-Geier, Lutz and Romanczuk, Pawel},
  journal={New J. Phys.},
  volume={15},
  number={8},
  pages={085014},
  year={2013},
  publisher={IOP Publishing}
}

@article{Barriere2015,
author = {Barriere, Guislaine and Fici, Pietro and Gallerani, Giulia and Fabbri, Francesco and Rigaud, Michel},
doi = {10.1186/s40169-015-0055-4},
issn = {2001-1326},
journal = {Clinic.Trans. Med.},
number = {1},
pages = {14},
title = {{Epithelial Mesenchymal Transition: a double-edged sword}},
url = {https://doi.org/10.1186/s40169-015-0055-4},
volume = {4},
year = {2015}
}

@article{Kohrman2017,
author = {Kohrman, Abraham Q. and Matus, David Q.},
doi = {10.1016/j.tcb.2016.08.003},
issn = {18793088},
journal = {T. Cell Bio.},
number = {1},
pages = {12--25},
pmid = {27634432},
publisher = {Elsevier Ltd},
title = {{Divide or Conquer: Cell Cycle Regulation of Invasive Behavior}},
url = {http://dx.doi.org/10.1016/j.tcb.2016.08.003},
volume = {27},
year = {2017}
}

@book{Bialek2012a,
author = {Bialek, William},
booktitle = {Biophys. Search. Princ.},
isbn = {9780691138916},
publisher = {Princeton University Press},
title = {{Biophysics: Searching for principles}},
year = {2012}
}

@article{Hoek2008,
author = {Hoek, Keith S. and Eichhoff, Ossia M. and Schlegel, Natalie C. and Dobbeling, U. and Kobert, Nikita and Schaerer, Leo and Hemmi, Silvio and Dummer, Reinhard},
doi = {10.1158/0008-5472.CAN-07-2491},
issn = {0008-5472},
journal = {Cancer Res.},
month = {2},
number = {3},
pages = {650--656},
title = {{In vivo switching of human melanoma cells between proliferative and invasive states}},
url = {http://cancerres.aacrjournals.org/cgi/doi/10.1158/0008-5472.CAN-07-2491},
volume = {68},
year = {2008}
}

@article{Prochazka2017,
author = {Prochazka, Laura and Benenson, Yaakov and Zandstra, Peter W.},
doi = {10.1016/j.coisb.2017.09.003},
issn = {24523100},
journal = {Curr. Opin. Syst. Biol.},
month = {10},
number = {September},
pages = {93--103},
publisher = {Elsevier Ltd},
title = {{Synthetic gene circuits and cellular decision-making in human pluripotent stem cells}},
url = {https://doi.org/10.1016/j.coisb.2017.09.003 https://linkinghub.elsevier.com/retrieve/pii/S2452310017301336},
volume = {5},
year = {2017}
}

@article{Jerby2012,
author = {Jerby, Livnat and Wolf, Lior and Denkert, Carsten and Stein, Gideon Y. and Hilvo, Mika and Oresic, Matej and Geiger, Tamar and Ruppin, Eytan},
doi = {10.1158/0008-5472.CAN-12-2215},
issn = {0008-5472},
journal = {Cancer Res.},
month = {11},
number = {22},
pages = {5712--5720},
title = {{Metabolic associations of reduced proliferation and oxidative stress in advanced breast cancer}},
url = {http://cancerres.aacrjournals.org/cgi/doi/10.1158/0008-5472.CAN-12-2215},
volume = {72},
year = {2012}
}

@article{Hatzikirou2018,
author = {Hatzikirou, Haralampos},
doi = {10.1515/jmbm-2018-0001},
issn = {2191-0243},
journal = {J. Mech. Behav. Mater.},
month = {4},
number = {1-2},
pages = {1--7},
title = {{Statistical mechanics of cell decision-making: the cell migration force distribution}},
url = {http://www.degruyter.com/view/j/jmbm.2018.27.issue-1-2/jmbm-2018-0001/jmbm-2018-0001.xml},
volume = {27},
year = {2018}
}

@article{Hatzikirou2010a,
author = {Hatzikirou, H. and Basanta, D. and Simon, M. and Schaller, K. and Deutsch, A.},
doi = {10.1093/imammb/dqq011},
issn = {1477-8599},
journal = {Math. Med. Biol.},
month = {3},
number = {1},
pages = {49--65},
title = {{'Go or Grow': the key to the emergence of invasion in tumour progression?}},
url = {https://academic.oup.com/imammb/article-lookup/doi/10.1093/imammb/dqq011},
volume = {29},
year = {2010}
}

@book{Murray2003,
address = {New York, NY},
author = {Murray, J. D.},
doi = {10.1007/b98869},
edition = {3},
editor = {Murray, J. D.},
isbn = {978-0-387-95228-4},
issn = {0939-6047},
pages = {1067},
publisher = {Springer New York},
series = {Interdisciplinary Applied Mathematics},
title = {{Mathematical Biology II. Spatial Models and Biomedical Applications}},
url = {http://www.springerlink.com/index/10.1007/978-0-387-75847-3 http://link.springer.com/10.1007/b98869},
volume = {18},
year = {2003}
}

@article{Bowsher2014,

author = {Bowsher, Clive G. and Swain, Peter S.},
doi = {10.1016/j.copbio.2014.04.010},
issn = {09581669},
journal = {Curr. Opin. Biotechnol.},
month = {8},
pages = {149--155},
publisher = {Elsevier Ltd},
title = {{Environmental sensing, information transfer, and cellular decision-making}},
url = {http://dx.doi.org/10.1016/j.copbio.2014.04.010 https://linkinghub.elsevier.com/retrieve/pii/S0958166914000822},
volume = {28},
year = {2014}
}

@article{Balazsi2011,
author = {Bal{\'{a}}zsi, G{\'{a}}bor and van Oudenaarden, Alexander and Collins, James J.},
doi = {10.1016/j.cell.2011.01.030},
issn = {00928674},
journal = {Cell},
month = {3},
number = {6},
pages = {910--925},
title = {{Cellular decision making and biological noise: from microbes to mammals}},
url = {https://linkinghub.elsevier.com/retrieve/pii/S0092867411000699},
volume = {144},
year = {2011}
}

@article{Price2003,
author = {Price, Trevor D. and Qvarnstr{\"{o}}m, Anna and Irwin, Darren E.},
doi = {10.1098/rspb.2003.2372},
issn = {0962-8452},
journal = {Proc. R. Soc. London. Ser. B Biol. Sci.},
month = {7},
number = {1523},
pages = {1433--1440},
pmid = {12965006},
title = {{The role of phenotypic plasticity in driving genetic evolution}},
url = {https://royalsocietypublishing.org/doi/10.1098/rspb.2003.2372},
volume = {270},
year = {2003}
}

@Article{kalluri,
  Title                    = {The basics of epithelial-mesenchymal transition},
  Author                   = {Raghu Kalluri AND Robert A. Weinberg},
  Journal                  = {The J. of Cli. Inves.},
  Year                     = {2010},

  Month                    = {5},
  Number                   = {5},
  Pages                    = {1786-1786},
  Volume                   = {120},

  Doi                      = {10.1172/JCI39104C1},
  Publisher                = {The American Society for Clinical Investigation},
  Url                      = {https://www.jci.org/articles/view/39104C1}
}

@Article{Radisky4325,
  Title                    = {Epithelial-mesenchymal transition},
  Author                   = {Radisky, Derek C.},
  Journal                  = {J. Cell Sci.},
  Year                     = {2005},
  Number                   = {19},
  Pages                    = {4325--4326},
  Volume                   = {118},

  Doi                      = {10.1242/jcs.02552},
  Eprint                   = {https://jcs.biologists.org/content/118/19/4325.full.pdf},
  ISSN                     = {0021-9533},
  Publisher                = {The Company of Biologists Ltd},
  Url                      = {https://jcs.biologists.org/content/118/19/4325}
}

@Article{THIERY2003740,
  Title                    = {Epithelial–mesenchymal transitions in development and pathologies},
  Author                   = {Jean Paul Thiery},
  Journal                  = {C. Opi. Cell Bio.},
  Year                     = {2003},
  Number                   = {6},
  Pages                    = {740 - 746},
  Volume                   = {15},
  Doi                      = {https://doi.org/10.1016/j.ceb.2003.10.006},
  ISSN                     = {0955-0674},
  Url                      = {http://www.sciencedirect.com/science/article/pii/S0955067403001339}
}

@Article{Hakim_2017,
  Title                    = {Collective cell migration: a physics perspective},
  Author                   = {Vincent Hakim and Pascal Silberzan},
  Journal                  = {Rep. on Prog. in Phys.},
  Year                     = {2017},

  Month                    = {4},
  Number                   = {7},
  Pages                    = {076601},
  Volume                   = {80},
  Doi                      = {10.1088/1361-6633/aa65ef},
  Publisher                = {{IOP} Publishing},
  Url                      = {https://doi.org/10.1088%2F1361-6633%2Faa65ef}
}

@Article{Schienbein1993,
  Title                    = {Langevin equation, Fokker-Planck equation and cell migration},
  Author                   = {Schienbein, M.
and Gruler, H.},
  Journal                  = {Bull. Math. Bio.},
  Year                     = {1993},


  Number                   = {3},
  Pages                    = {585--608},
  Volume                   = {55},
  Day                      = {01},
  Doi                      = {10.1007/BF02460652},
  ISSN                     = {1522-9602},
  Url                      = {https://doi.org/10.1007/BF02460652}
}

@Article{Milster2017,
  Title                    = {Eliminating inertia in a stochastic model of a micro-swimmer with constant speed},
  Author                   = {Milster, S.
and N{\"o}tel, J.
and Sokolov, I. M.
and Schimansky-Geier, L.},
  Journal                  = {The Euro. Phys. Jour. Sp. Topics},
  Year                     = {2017},


  Number                   = {9},
  Pages                    = {2039--2055},
  Volume                   = {226},
  Day                      = {01},
  Doi                      = {10.1140/epjst/e2017-70052-8},
  ISSN                     = {1951-6401},
  Url                      = {https://doi.org/10.1140/epjst/e2017-70052-8}
}

@article{bhatt,
  title = {A Model for Collision Processes in Gases. I. Small Amplitude Processes in Charged and Neutral One-Component Systems},
  author = {Bhatnagar, P. L. and Gross, E. P. and Krook, M.},
  journal = {Phys. Rev.},
  volume = {94},
  issue = {3},
  pages = {511--525},
  numpages = {0},
  year = {1954},

  publisher = {American Physical Society},
  doi = {10.1103/PhysRev.94.511},
  url = {https://link.aps.org/doi/10.1103/PhysRev.94.511}
}

@article{dahlman,
author = {Dahlman, E and Watanabe, Y},
title = {SU-F-T-109: A Shortcoming of the Fisher-Kolmogorov Reaction-Diffusion Equation for Modeling Tumor Growth},
journal = {Med. Phys.},
volume = {43},
number = {6Part14},
pages = {3486-3487},
doi = {10.1118/1.4956245},
url = {https://aapm.onlinelibrary.wiley.com/doi/abs/10.1118/1.4956245},
eprint = {https://aapm.onlinelibrary.wiley.com/doi/pdf/10.1118/1.4956245},
year = {2016}
}

@article{banyard,
author = {Jacqueline Banyard and Diane R. Bielenberg},
title = {The role of EMT and MET in cancer dissemination},
journal = {Conn. Tis. Res.},
volume = {56},
number = {5},
pages = {403-413},
year  = {2015},
publisher = {Taylor & Francis},
doi = {10.3109/03008207.2015.1060970},
    note ={PMID: 26291767},

URL = { 
        https://doi.org/10.3109/03008207.2015.1060970
    
},
eprint = { 
        https://doi.org/10.3109/03008207.2015.1060970
    
}

}

@Article{Malinverno2017,
author={Malinverno, Chiara
and Corallino, Salvatore
and Giavazzi, Fabio
and Bergert, Martin
and Li, Qingsen
and Leoni, Marco
and Disanza, Andrea
and Frittoli, Emanuela
and Oldani, Amanda
and Martini, Emanuele
and Lendenmann, Tobias
and Deflorian, Gianluca
and Beznoussenko, Galina V.
and Poulikakos, Dimos
and Ong, Kok Haur
and Uroz, Marina
and Trepat, Xavier
and Parazzoli, Dario
and Maiuri, Paolo
and Yu, Weimiao
and Ferrari, Aldo
and Cerbino, Roberto
and Scita, Giorgio},
title={Endocytic reawakening of motility in jammed epithelia},
journal={Nature Mat.},
year={2017},
volume={16},
number={5},
pages={587-596},
issn={1476-4660},
doi={10.1038/nmat4848},
url={https://doi.org/10.1038/nmat4848}
}

@article{Yan,
  title = {Multicellular Rosettes Drive Fluid-solid Transition in Epithelial Tissues},
  author = {Yan, Le and Bi, Dapeng},
  journal = {Phys. Rev. X},
  volume = {9},
  issue = {1},
  pages = {011029},
  numpages = {18},
  year = {2019},

  publisher = {American Physical Society},
  doi = {10.1103/PhysRevX.9.011029},
  url = {https://link.aps.org/doi/10.1103/PhysRevX.9.011029}
}

@Article{Hanahan2000,
author={Hanahan, Douglas
and Weinberg, Robert A.},
title={The Hallmarks of Cancer},
journal={Cell},
year={2000},

day={07},
publisher={Elsevier},
volume={100},
number={1},
pages={57-70},
issn={0092-8674},
doi={10.1016/S0092-8674(00)81683-9},
url={https://doi.org/10.1016/S0092-8674(00)81683-9}
}

@book{bruce,
 author = "Alberts, B.",
 title = "Molecular Biology of the Cell",
 publisher = "New York: W. W",
 year = 2015
}

@book{box,
  title={Statistics for Experimenters: An Introduction to Design, Data Analysis, and Model Building},
  author={George E. P. Box and William G. Hunter and Joanna S. Hunter},
  year={1979}
}

@Article{Jollyrsc,
  Title                    = {Towards elucidating the connection between epithelial - mesenchymal transitions and stemness},
  Author                   = {Jolly, Mohit Kumar and Huang, Bin and Lu, Mingyang and Mani, Sendurai A. and Levine, Herbert and Ben-Jacob, Eshel},
  Journal                  = {J. Roy. Soc. Int.},
  Year                     = {2014},
  Number                   = {101},
  Pages                    = {20140962},
  Volume                   = {11},
  Doi                      = {10.1098/rsif.2014.0962},
  Eprint                   = {https://royalsocietypublishing.org/doi/pdf/10.1098/rsif.2014.0962},
  Url                      = {https://royalsocietypublishing.org/doi/abs/10.1098/rsif.2014.0962}
}

@Article{BoaretoE402,
  Title                    = {Jagged{\textendash}Delta asymmetry in Notch signaling can give rise to a Sender/Receiver hybrid phenotype},
  Author                   = {Boareto, Marcelo and Jolly, Mohit Kumar and Lu, Mingyang and Onuchic, Jos{\'e} N. and Clementi, Cecilia and Ben-Jacob, Eshel},
  Journal                  = {PNAS},
  Year                     = {2015},
  Number                   = {5},
  Pages                    = {E402--E409},
  Volume                   = {112},
  Doi                      = {10.1073/pnas.1416287112},
  Eprint                   = {https://www.pnas.org/content/112/5/E402.full.pdf},
  ISSN                     = {0027-8424},
  Publisher                = {National Academy of Sciences},
  Url                      = {https://www.pnas.org/content/112/5/E402}
}

@Article{Barua404889,
  Title                    = {A least microenvironmental uncertainty principle (LEUP) as a generative model of collective cell migration mechanisms},
  Author                   = {Barua, Arnab and Nava-Sede{\~n}o, Josue M. and Hatzikirou, Haralampos},
  Journal                  = {bioRxiv},
  Year                     = {2019},
  Doi                      = {10.1101/404889},
  Elocation-id             = {404889},
  Eprint                   = {https://www.biorxiv.org/content/early/2019/10/04/404889.full.pdf},
  Publisher                = {Cold Spring Harbor Laboratory},
  Url                      = {https://www.biorxiv.org/content/early/2019/10/04/404889}
}
\newpage
\section*{Supplementary Information}
\begin{center}
\begin{tabular}{ |c|c| } 
 \hline
Symbol & Explanation \\ \hline
 $X_{i}$ & Phenotype of $i$-th cell\\
 $Y_{i}$ & Phenotype of $i$-th neighbourhood cells \\
 $N_{i}^{0}$ & Number of cells in $i$-th cell's microenvironment having phenotype $(X_{i}=0)$ \\ 
 $N_{i}^{1}$ & Number of cells in $i$-th cell's microenvironment having phenotype $(X_{i}=1)$ \\ 
 $N_{i}^{\phi}$ & Number of free slots in $i$-th cell's microenvironment \\ 
 $V$ & Total capacity \\ 
  $N_{T}$ & Total number of phenotypes\\ 
     $\ell$ & Radius of the microenvironment \\ 
      $\nu$ & Exchange rate\\ 
      $r$ & Growth rate\\
       $\beta$ & LEUP sensitivity\\
       $\rho_{0}(x)$ & Mean density of the resting cells at position x \\
    $\rho_{1}(x)$ & Mean density of the migratory cells at position x \\
     $D_0$ & Diffusion coefficient of resting cells\\
      $D_1$ & Diffusion coefficient of migratory cells\\
 \hline
\end{tabular}
\end{center}
\section*{S.1. Calculation of microenvironmental entropy}
We assume that a maximum number of N cells is present inside the cell's
microenvironment, where l is the radius of the microenvironment (Fig.\ref{scheme1}). Since
there is the possibility of free slots, if there are less than N cells in the cell
neighbourhood, we have started by assuming a trinomial distribution for the $i$-th cell is $P\left(N^{1}_{i},N^{\phi}_{i}\right)$ where the number $ N_{i}^{0} $ of cells having phenotype $X_{i} = 0$ and by the number $ N_{i}^{1} $ of cells having phenotype $X_{i} = 1$. In addition, a number $
N_{i}^{\phi}$ of empty slots are included inside the microenvironment. The joint probability is defined by 
\begin{equation}
    P\left(N^{1}_{i},N^{\phi}_{i}\right)=\frac{N!}{N^{1}_{i}!N^{\phi}_{i}!\left(N-N^{\phi}_{i}-N^{1}_{i}\right)!}p^{N^{1}_{i}}\theta^{N^{\phi}_{i}}\left(1-p-\theta\right)^{\left(N-N^{\phi}_{i}-N^{1}_{i}\right)},
\end{equation}
where $p$ and $\theta$ are the probabilities of having a number $N_i^{1}$ of cells with phenotype
$( X_{i} = 1)$ out of $N$ cells and having a number $N_i^{\phi}$ of free slots. The conditional
probability of having a number $N_i^{1}$ of cells present in the microenvironment given a
number $N_i^{\phi}$ of free slots is
\begin{equation}
\begin{split}
   & P\left(N_{i}^{1}\mid N_{i}^{\phi} \right)=\frac{ P\left(N_{i}^{1}, N_{i}^{\phi} \right)}{ P\left( N_{i}^{\phi} \right)}\\
    &=\frac{\frac{N!}{N^{1}_{i}!N^{\phi}_{i}!\left(N-N^{\phi}_{i}-N^{1}_{i}\right)!}p^{N^{1}_{i}}\theta^{N^{\phi}_{i}}\left(1-p-\theta\right)^{\left(N-N^{\phi}_{i}-N^{1}_{i}\right)}}{\frac{N!}{N^{\phi}_{i}!\left(N-N^{\phi}_{i}\right)!}\theta^{N^{\phi}_{i}}\left(1-\theta\right)^{\left(N-N^{\phi}_{i}\right)}}\\
    &=\binom{N-N^{\phi}_{i}}{N^{1}_{i}}\left(\frac{p}{1-\theta}\right)^{N^{1}_{i}}\left(1-\frac{p}{1-\theta}\right)^{N-N^{1}_{i}-N^{\phi}_{i}}\\
    &= \boldsymbol{B}\left(N- N_{i}^{\phi}, \frac{p}{1-\theta} \right)\\
    &= \boldsymbol{B}\left(N_{T}, \frac{p}{1-\theta} \right)
    \end{split}
    \end{equation}
    Now, we use the form of binomial distribution to calculate the entropy of the Binomial distribution. Please note that $P(N_i^1\mid N_i^\phi)$ can be written as $P(N_i^1\mid N_T)$ due to Binomial distribution. 
    \begin{equation}
          S\left(N^{1}_{i}\mid N_{T}\right)=S\left(Y_{i}\right)=\frac{1}{2}\log_{2}\left(2\pi e\frac{N_{T}p}{1-\theta}\left(1-\frac{p}{1-\theta}\right)\right),  
    \end{equation}
where
\begin{equation}
    \begin{split}
        &\frac{p}{1-\theta}=\frac{N^{1}_{i}}{N}\frac{1}{1-\frac{N^{\phi}_{i}}{N}}=\frac{N^{1}_{i}}{N-N^{\phi}_{i}}=\frac{N_{i}^{1}}{N_{T}}, \\
        &1-\frac{p}{1-\theta}=\frac{N^{0}_{i}}{N-N^{\phi}_{i}}=\frac{N_{i}^{0}}{N_{T}}. 
    \end{split}
\end{equation}
According to LEUP we have to evaluate the microenvironmental  entropy of $S\left(Y_i\mid X_i =0\right)$ and $S\left(Y_i\mid X_i = 1\right)$ to calculate the probability of the internal states $\left(X_{i}=0\right)$ and $\left(X_{i}=1\right)$ 
\begin{equation}
    \begin{split}
&    S\left(Y_i\mid X_i =1\right)=\frac{1}{2}\log_{2}\left(2\pi e\left(\frac{\left(N_{T}-1\right) \left(N^{1}_{i}-1\right)}{\left(N_{T}-1\right)}\right)\left(\frac{N^{0}_{i}}{\left(N_{T}-1\right)}\right)\right),\\
&  S\left(Y_i\mid X_i =0\right)=\frac{1}{2}\log_{2}\left(2\pi e\left(\frac{\left(N_{T}-1\right) \left(N^{1}_{i}\right)}{\left(N_{T}-1\right)}\right)\left(\frac{N^{0}_{i}-1}{\left(N_{T}-1\right)}\right)\right).
\end{split}
\end{equation}
From the Gaussian approximation we can write the entropy difference as
\begin{equation}
\begin{split}
&\Delta S=S(Y_{i}|X_{i} =1)- S(Y_{i}| X_{i} = 0),\\
&=\frac{1}{2}\ln{\left[\frac{N_{i}^{0}\left(N_{i}^{1}-1\right)}{N_{i}^{1}\left(N_{i}^{0}-1\right)}\right]}.\\
\end{split}
\end{equation}

\section*{S.2. Calculation of cell proliferation rate}
In our go-or-grow model only resting cells are allowed to proliferate but not the
moving cells. The growth rate depends in a specific way on the number of moving
and resting cells in the microenvironment. In particular, we assume that the per-
capita growth rate for resting cells is a linearly decreasing function of $N_i^0$ and $N_i^1$,
and is also decreasing with
the number of migratory cells and a constant per-capita death rate $d_1$. Accordingly,

\begin{equation}
\begin{aligned}
    &\frac{d \langle N_{i}^{0}\rangle}{d t}=\left(\langle W^{+}\rangle_{N_{i}^{0},N_{i}^{1}}-\langle W^{-}\rangle_{N_{i}^{0},N_{i}^{1}}\right)\\
    &=h_{1}\rho_{0}-q\left(\rho_{0}+\rho_{1}\right)\rho_{0}-d_{1}\rho_{0}\\
    &=r\rho_{0}\left(1-\rho_{0}-\rho_{1}\right)
    \end{aligned}
\label{prolif}
\end{equation}
where $ \frac{q}{r} =1 $ and $r=h_{1}-d_{1}$.
 \newpage
 \begin{figure}
 \section*{S.4. Phase space diagram for null sensitivity case}
 \centering
\includegraphics[scale=1]{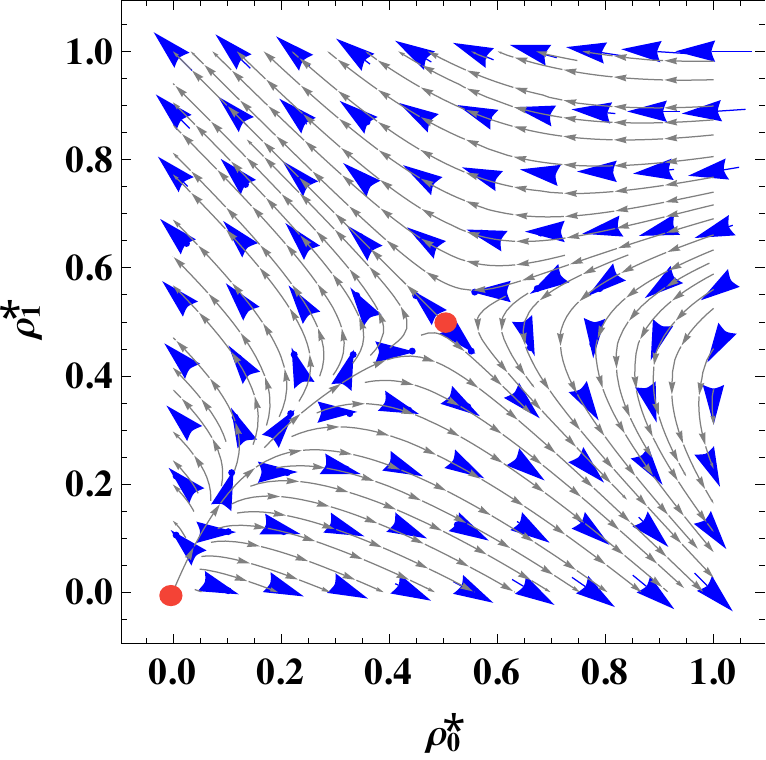}
\caption{Phase space diagram of $\rho_{0}^{*}$ and $ \rho_{1}^{*}$ for $r= 1$} where two fixed points are marked by red circles.
\label{phase}
\end{figure}
 \begin{figure}
 \section*{S.5. Turing space for pattern formation}
    \centering
  \begin{subfigure}[b]{0.45\linewidth}
\includegraphics[width=\linewidth]{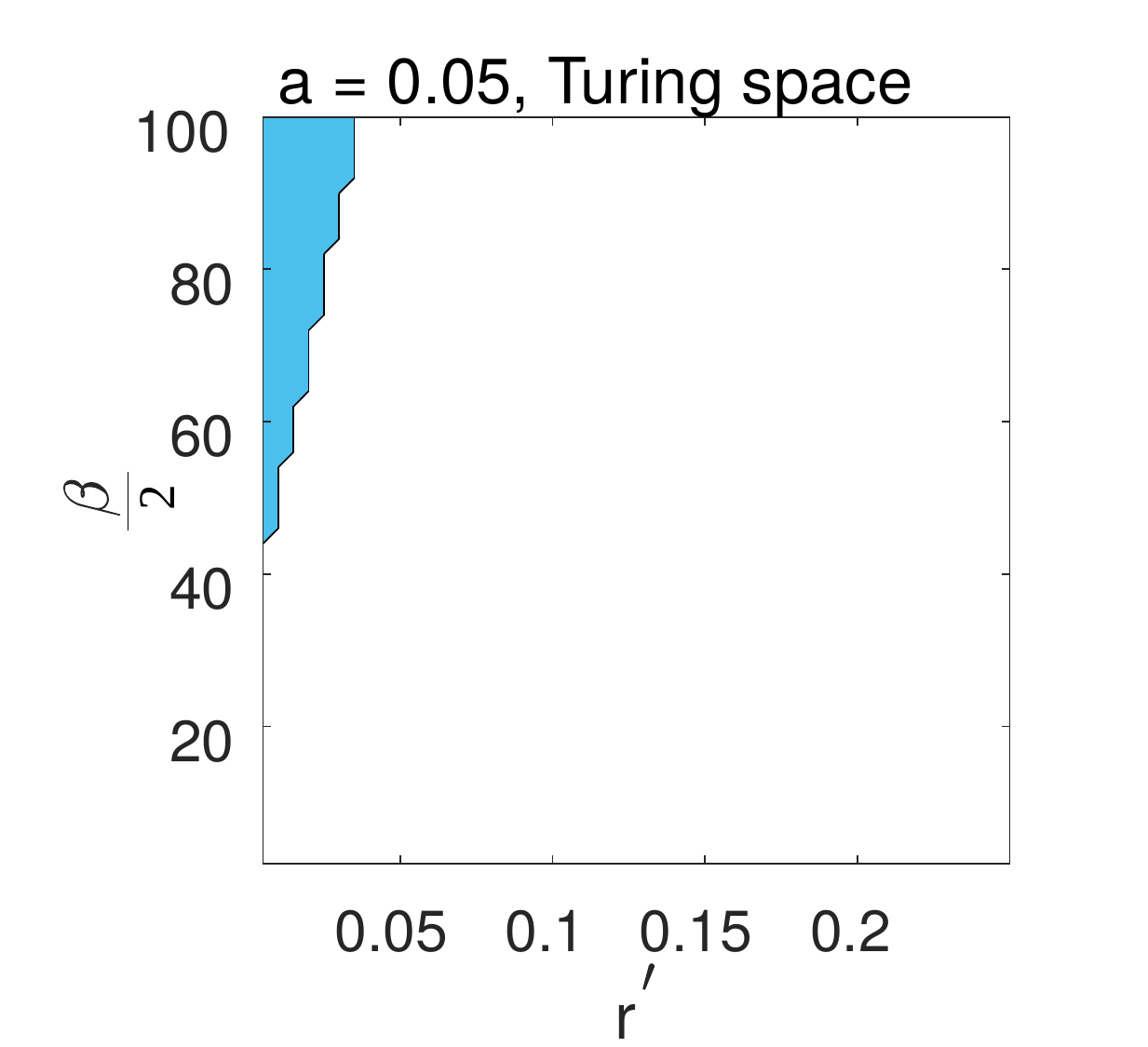}
\caption{$a$ is fixed at 0.05.}\label{turing0.05}
\end{subfigure}
~
\begin{subfigure}[b]{0.45\linewidth}
\includegraphics[width=\linewidth]{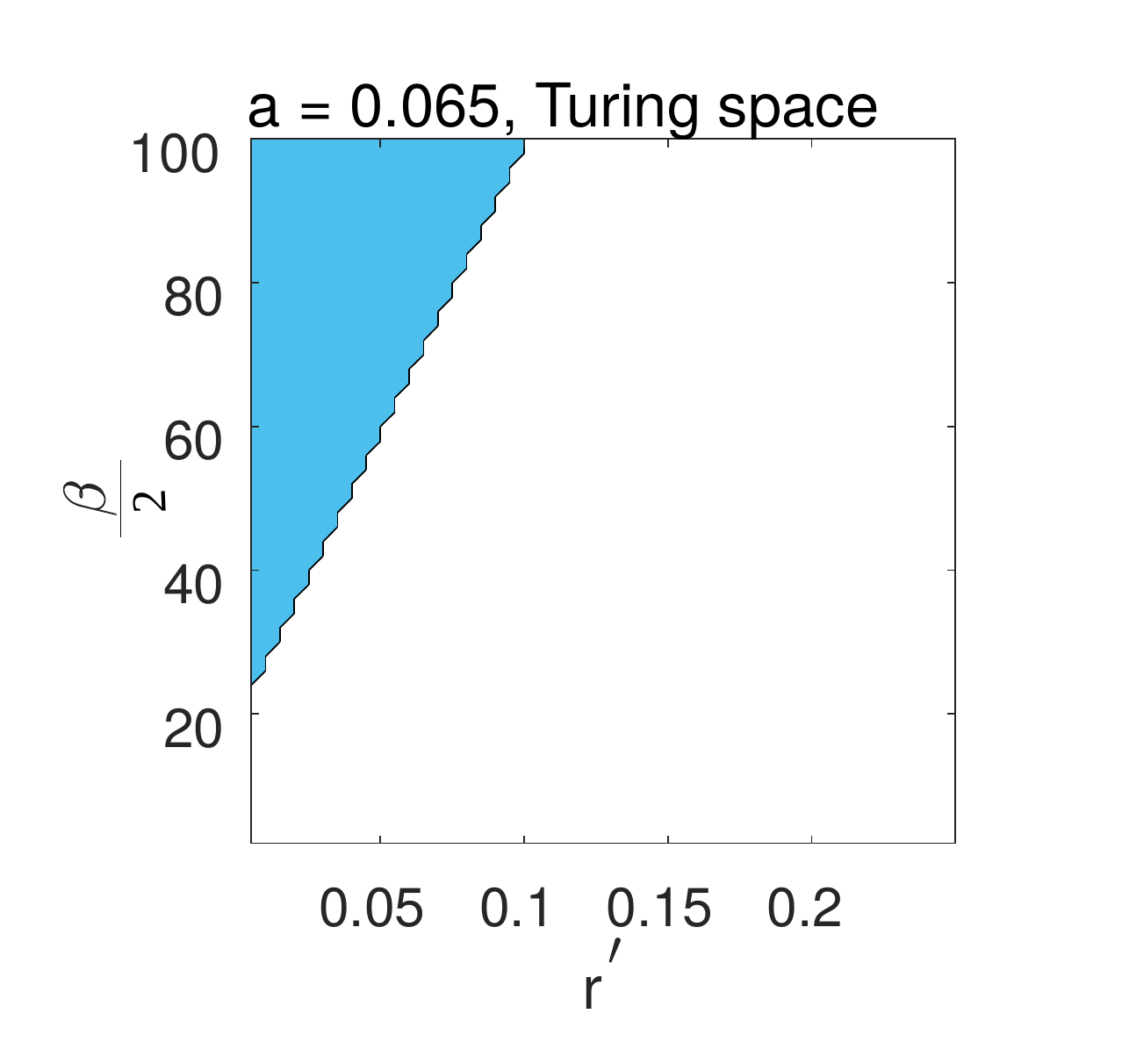}
\caption{$a$ is fixed at 0.065.}\label{turng0.065}
\end{subfigure}
~
\begin{subfigure}[b]{0.45\linewidth}
\includegraphics[width=\linewidth]{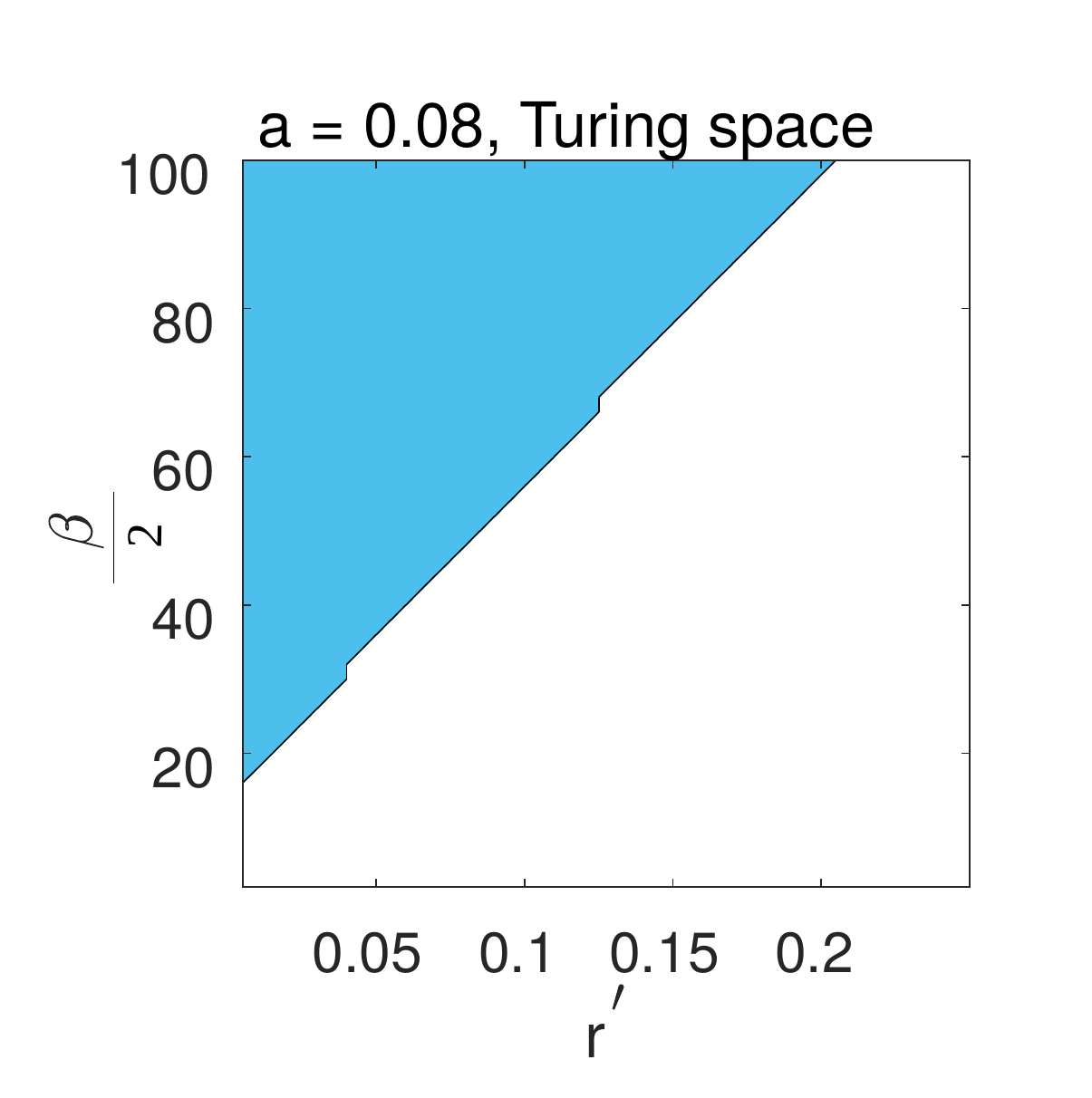}
\caption{$a$ is fixed at 0.08.}\label{turng0.08}
\end{subfigure}
\caption{Phase space diagram of $\beta$ vs. $r^{'}$ for different values of $a$ (\textbf{a-c}), where the Turing
space is marked in blue. $a$ is defined by $\frac{1}{V}$. }
\label{turing}
\end{figure}
\begin{figure}
\section*{S.6. Critical wavelengths inside the Turing space for pattern formation}
    \centering
  \begin{subfigure}[b]{0.45\linewidth}
\includegraphics[width=\linewidth]{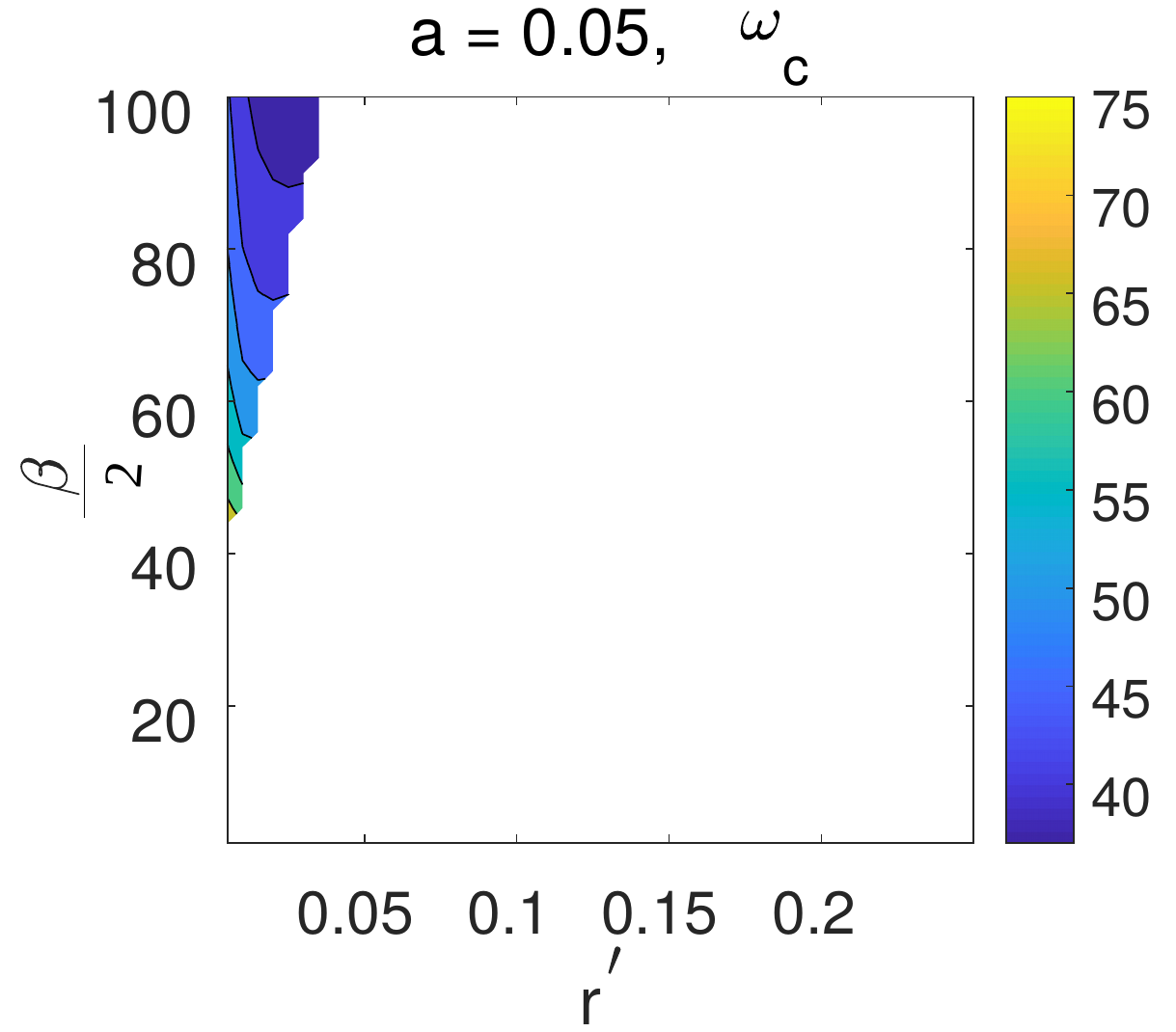}
\caption{$a$ is fixed at 0.05.}\label{turing0.01wc}
\end{subfigure}
~
\begin{subfigure}[b]{0.45\linewidth}
\includegraphics[width=\linewidth]{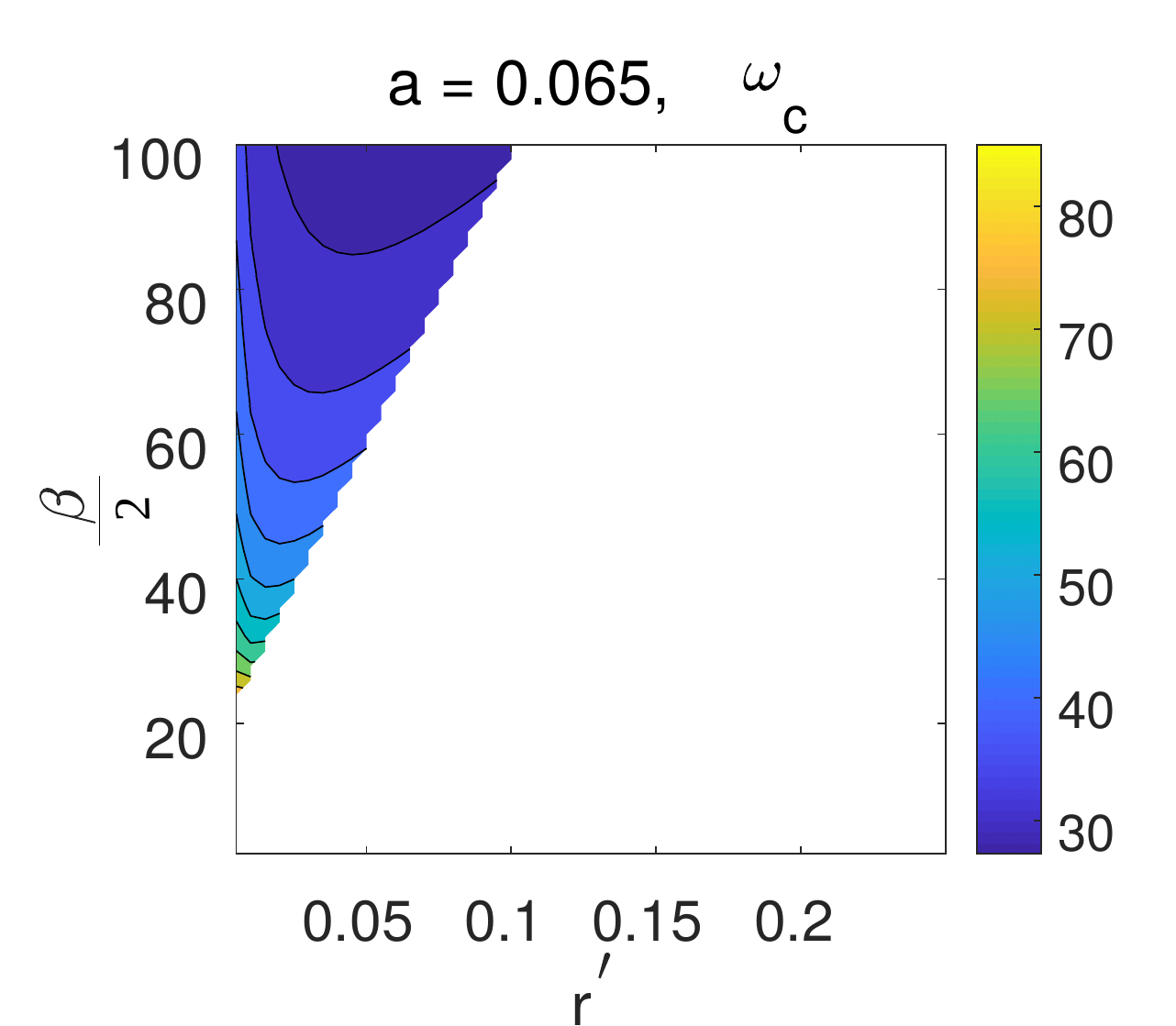}
\caption{$a$ is fixed at 0.065.}\label{turng0.05wc}
\end{subfigure}
~
\begin{subfigure}[b]{0.45\linewidth}
\includegraphics[width=\linewidth]{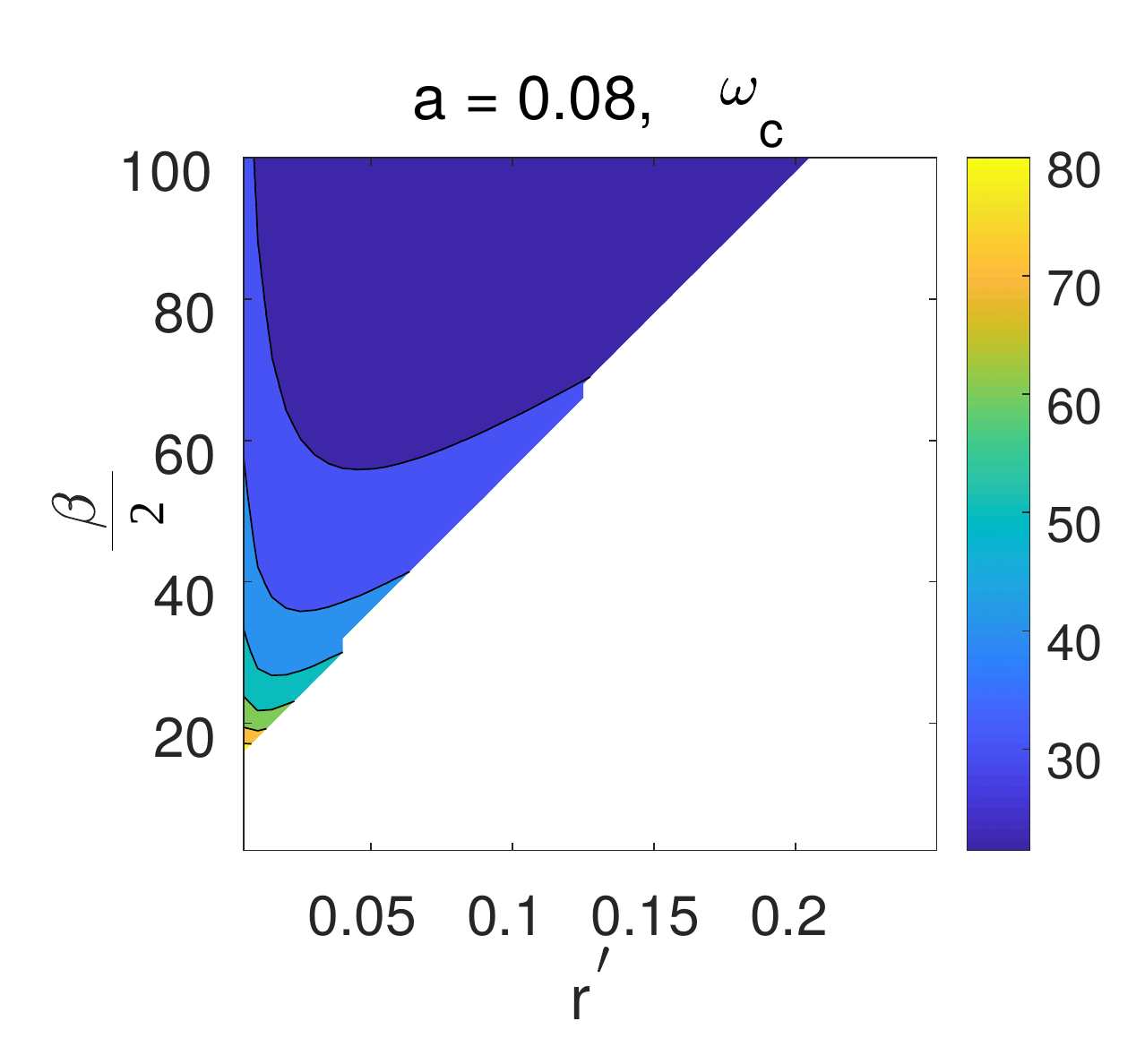}
\caption{$a$ is fixed at 0.08.}\label{turng0.03wc}
\end{subfigure}
\caption{Phase space diagram of $\beta$ vs. $r^{'}$ for different values of $a$ (\textbf{a-c}), where critical wavelengths $(\omega_c)$ have been plotted inside the Turing space. $a$ is defined by $\frac{1}{V}$.}
\end{figure}
\label{turingwc}
\begin{figure}
\section*{S.7. Plot of resting probability}
    \centering
    \includegraphics[scale=0.8]{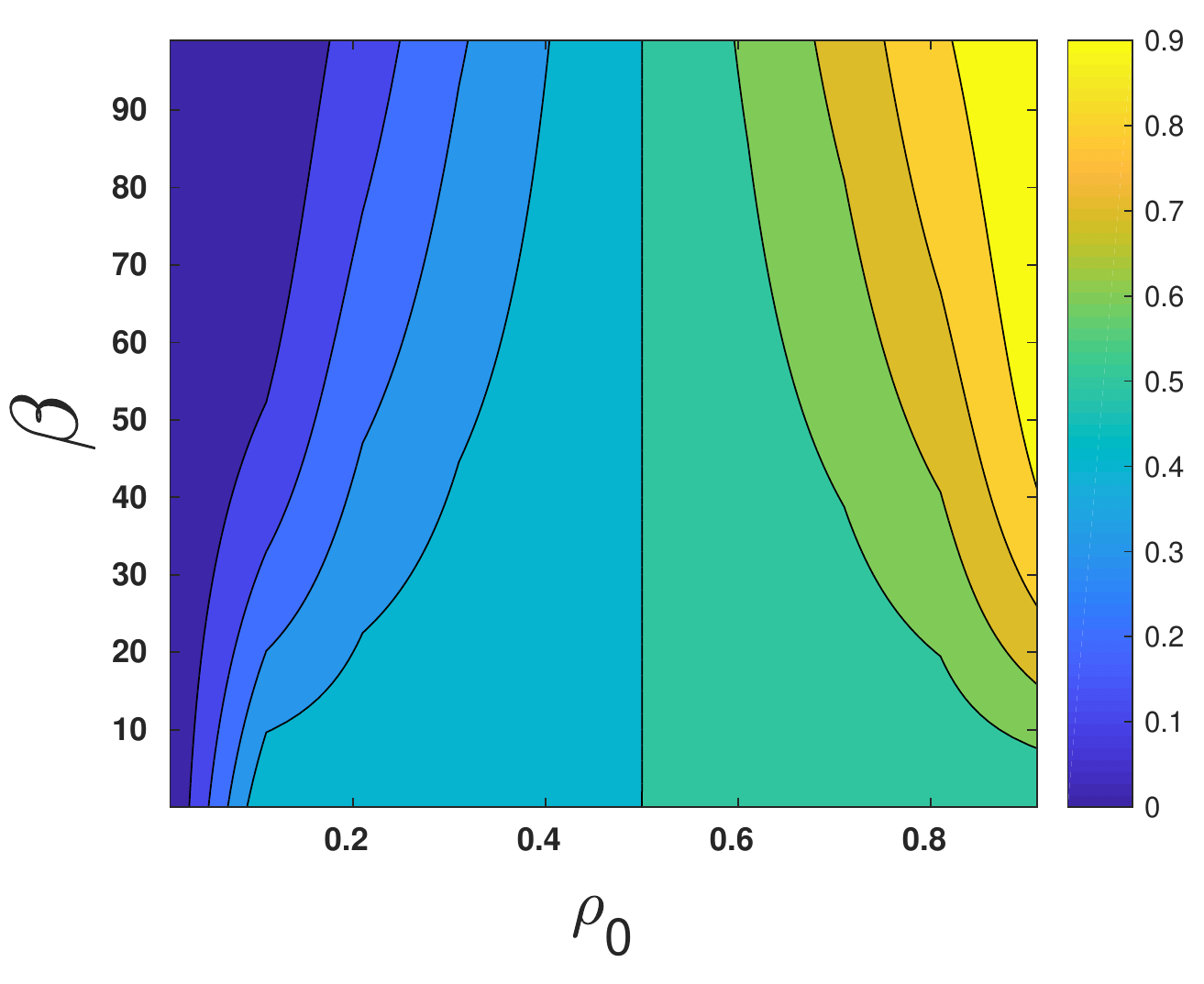}
    \caption{Plot of resting probability with respect to $\beta$ and $\rho_0$. We have fixed the value of $\frac{1}{V}$ to 0.01.}
    \label{restprob}
\end{figure}
\begin{figure}

\section*{S.8. Critical sensitivity vs. inverse capacity graph from IBM}
    \centering
    \includegraphics[scale=0.8]{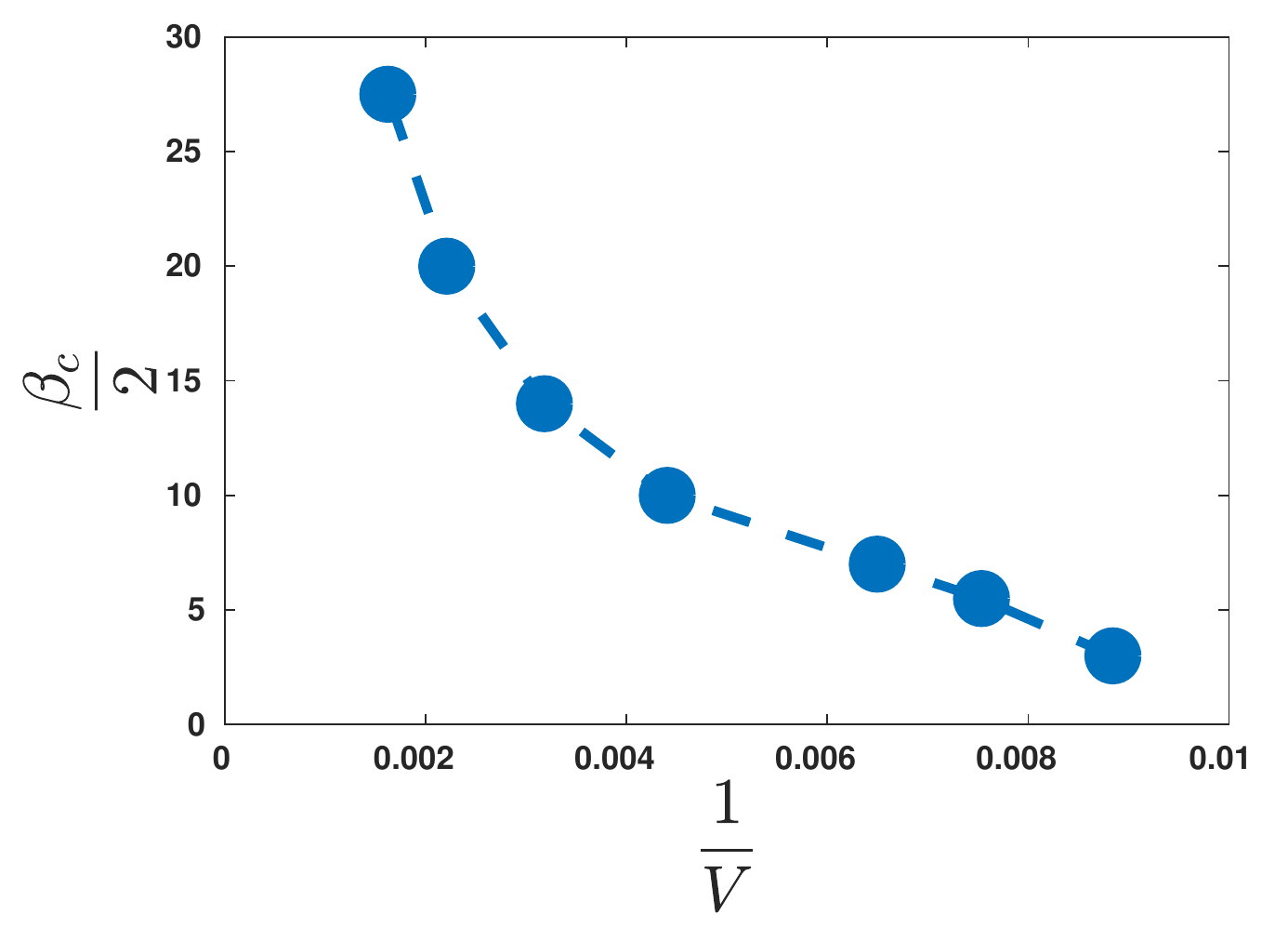}
    \caption{Critical sensitivity with respect to the inverse capacity of the IBM over 5 simulations. Throughout the simulations, we kept the total density at 0.2 and the total number of cells was 500.}
    \label{critibm}
\end{figure}
\end{document}